\documentclass[12pt,a4paper,twoside]{report}

\usepackage[utf8]{inputenc}
\usepackage[T1]{fontenc}
\usepackage{palatino}
\usepackage{inconsolata}
\usepackage{mathpazo}

\usepackage{subfiles}
\usepackage{fullpage}

\usepackage{color}
\usepackage[dvipsnames]{xcolor}
\usepackage{tikz}
\usetikzlibrary{shapes.misc}
\tikzset{cross/.style={cross out, draw=black, minimum size=2*(#1-\pgflinewidth), inner sep=0pt, outer sep=0pt}, cross/.default={1pt}}
\usepackage{pgfplots}
\usepgfplotslibrary{fillbetween}
\usetikzlibrary{patterns}

\usepackage{framed}
\usepackage{booktabs}
\usepackage{tabularx}
\usepackage{multirow}

\usepackage{listings}
\usepackage{amsmath}
\usepackage{amssymb}

\usepackage{xfrac}
\usepackage{mathtools}

\DeclarePairedDelimiter{\floor}{\lfloor}{\rfloor}

\usepackage[font=small,labelfont=bf,margin=6pt]{caption}
\usepackage[hidelinks]{hyperref}

\hypersetup{pdftitle={Time-tiling in Devito},pdfauthor={Nicholas Sim}}

\newcommand{\bb}[0]{\mathcal{B}}

\usepackage{afterpage}

\newcommand\blankpage{%
	\null
	\thispagestyle{empty}%
	\addtocounter{page}{-1}%
	\newpage}

\begin{document}
	\begin{titlepage} 
	\newcommand{\HRule}{\rule{\linewidth}{0.5mm}} 

	\centering 


	\textsc{\LARGE Imperial College of Science, Technology and Medicine}\\[1.5cm]

	\textsc{\large Department of Computing}\\[0.5cm] 
	
	\textsc{\large BEng Mathematics and Computer Science Individual Project}\\[0.5cm] 
	
	
	\HRule\\[0.4cm]
	
	{\huge\bfseries Optimising finite-difference methods for PDEs through parameterised time-tiling in Devito}\\[0.15cm] 
	
	\HRule\\[1.5cm]
	
	
	\begin{minipage}{0.4\textwidth}
		\begin{flushleft}
			\large
			Nicholas \textsc{Sim}\\~\\~\\~\\~
		\end{flushleft}
	\end{minipage}
	~
	\begin{minipage}{0.4\textwidth}
		\begin{flushright}
			\large
			\textit{Supervisors}\\
			Prof.~Paul H. J. \textsc{Kelly}\\
			Dr~Fabio \textsc{Luporini}\\
			\textsc{Sun} Tianjiao\\
			~\\
			\textit{Second marker}\\
			Dr~Gerard \textsc{Gorman}
		\end{flushright}
	\end{minipage}



	\vfill\vfill\vfill 

	{\large June 2018} 



	\vfill 

\end{titlepage}
	\afterpage{\blankpage}

	\setlength{\parskip}{0.6em plus0.25em minus0.2em}
	\addcontentsline{toc}{chapter}{Abstract}

\begin{abstract}
\setcounter{page}{3}

Finite-difference methods are widely used in solving partial differential equations.
In a large problem set, approximations can take days or weeks to evaluate, yet the bulk of computation may occur within a single loop nest.
The modelling process for researchers is not straightforward either, requiring models with differential equations to be translated into stencil kernels, then optimised separately.
One tool that seeks to speed up and eliminate mistakes from this tedious procedure is Devito, used to efficiently employ finite-difference methods.

In this work, we implement \emph{time-tiling}, a loop nest optimisation, in Devito yielding a decrease in runtime of up to 45\%, and at least 20\% across stencils from the acoustic wave equation family, widely used in Devito's target domain of seismic imaging.
We present an estimator for \emph{arithmetic intensity under time-tiling} and a model to predict runtime improvements in stencil computations. We also consider generalisation of time-tiling to imperfect loop nests, a less widely studied problem.
\end{abstract}
\afterpage{\blankpage}
	\renewcommand{\abstractname}{Acknowledgements}
\addcontentsline{toc}{chapter}{Acknowledgements}

\begin{abstract}
	\setcounter{page}{5}

	To my supervisors, Prof.~Paul Kelly and Dr.~Fabio Luporini, my deepest thanks are due for their invaluable advice and insight, for the clearest explanations which I can only hope to emulate, and for igniting my interest in computer architecture.
	I must also thank Sun Tianjiao, who provided much-needed early guidance on performance testing and regular advice throughout.
	These people made this project a delight to work on.

	Thanks also to my parents, for their boundless support. To them I owe this and my sense of determination. Finally, my unqualified gratitude goes to friends who remind me of the greater context of this---and every---work.
\end{abstract}

\afterpage{\blankpage}

	\setcounter{page}{7} 
	\setlength{\parskip}{0.0pt plus 1.0pt}
	\tableofcontents{}

	\setlength{\parskip}{0.6em plus0.25em minus0.2em}
	\documentclass[thesis.tex]{subfile}

\chapter{Introduction}

\section{Motivation}

In many engineering domains, finite-difference methods are used to find approximate solutions to differential equations.
For scientists using differential equations in novel ways, tools which can transform partial differential equations to optimised computations and evaluate them are useful; this enables them to refine their models quickly.

It is inefficient and undesirable for specialists principally concerned with modelling to optimise their computations by hand.
As a means of abstraction, compilers can be used to automatically transform differential equations into stencils, then code to evaluate the equations.
A tool automatically performing this transformation saves considerable labour and scope for errors, and reduces the effort required to comprehend the calculations.
These aid maintenance and even reproducibility; generating and optimising code by hand is rarely feasible or efficient~\cite{olgaard10}.

Decoupling optimisation from specific models means that domain specialists can take advantage of optimisations seamlessly without intervention, allowing them to focus their attention and expertise on their own domain.
They can easily take advantages of the latest advances in computer architecture and new instructions such as vectorisation, which give rise to new optimisations in finite-difference methods.

\subsection*{Devito: a tool for solving differential equations}

In modern usage, the partial differential equations are transformed into \emph{stencils}, which define the computation, then code to execute it.
One of the challenges faced by domain specialists generating new models is the time necessary to perform this translation.
Thus the desire for a compiler to save time and effort arises.

Devito~\cite{devito}, a tool for efficient application of finite-difference methods, is able to generate computations directly from differential equations, achieving a notion of `vertical integration' within the modelling ecosystem.
There are several compilers, a number of which we review in Chapter~\ref{ch:background}, which perform optimisation on stencil kernels, but none which also transform differential equations into stencils.
Being able to automate this transformation is extremely helpful if one is experimenting with models, or continually modelling new problems, as one can change the equations used and rapidly generate and execute the relevant computation.

\subsection*{Context and existing work in time-tiling}
Some stencil compilers are able to apply the \emph{time-tiling} optimisation, which this work implements in Devito.
It was previously demonstrated that time-tiling could reduce the run time of `some Devito stencil loops by up to 27.5\%'~\cite{dylan}, comparing time-tiling against the existing spatial tiling which the tool is able to perform.
In this case another tool, CLooG~\cite{cloog-isl}, was used to add time-tiling to Devito-generated code.

While the literature surrounding the polyhedral model, from which time-tiling derives, dates from the last century, it was viewed as complicated and time-consuming to use in optimising compilers as recently as 2004 when Bastoul proposed extensions~\cite{cloog-isl} to the original Quiller\'e et al.~algorithm to eliminate redundant (generated) code~\cite{quillere}.
Time-tiling has been studied in the form this work considers since Griebl in 2004~\cite{griebl-tt}, using \emph{space-time mapping}, analogous to the skewed tiling which we discuss in this work, but expressed as affine conditions specifying hyperplane partitions of the problem domain.

Time-tiling has since been studied extensively, and long implemented in polyhedral compilers such as PLUTO, providing a strong theoretical foundation.
However, until this work, the actual application of the technique beyond theoretical study has been limited to the OPS (Oxford Parallel library for Structured mesh solvers) project~\cite{ops-main}, who presented their implementation of time-tiling in 2017~\cite{ops-tiling}.
Even so, OPS is a source-to-source compiler, operating from the stencil kernel level down to code generation; Devito includes all these plus layers of abstraction up to the differential equations themselves, making this a unique implementation and study of time-tiling.

\subsection*{Objective}
Therefore, we extend tiling to the time dimension natively in Devito to realise this performance gain and evaluate its efficacy.
This gives the performance gains of time-tiling to Devito without the need to configure another tool to perform further optimisations.

\section{Contributions}
The principal objective of this project was to implement tiling over the time dimension in Devito and evaluate its performance against the existing optimisation, that of tiling restricted to the spatial dimensions.

In summary, the contributions of this thesis are:

\begin{itemize}
	\item A fully-functional implementation of time-tiling for perfect loop nests in Devito, including a simplification of the generated code structure, with accompanying test cases and auto-tuner enhancements.
	An analysis of its legality and the necessary conditions to guarantee this.

	\item An evaluation of the correctness and performance of the time-tiling transformation and any actions end-users may need to take to realise performance gains.

	\item Demonstration of a runtime decrease of up to 45\%, and in general more than 20\%, compared to Devito's existing optimisations on kernels from real-world applications, including stencils from the family of acoustic wave equation operators, widely used in Devito's target domain, seismic imaging.

	\item A novel estimator for \emph{arithmetic intensity under time-tiling} that extends the existing estimator for arithmetic intensity under spatial tiling in Devito, and a proof of its consistency with the widely-cited roofline model.
	An evaluation of the utility of this estimator compared to a generalised version of the existing estimator, and an analysis of the shortcomings of both.
	A model to predict when performance improvement is minimal or negative, and a demonstration that stencils of the acoustic wave equation family can benefit further from greater cache reuse.

	\item An analysis on how parameters introduced by time-tiling, the time tile size and skewing factor, affect the runtime of stencil computations, and the circumstances under which the runtime decrease is maximal.

	\item A discussion on further implementation work, analysing their importance and consistency with the transformation we have implemented.
\end{itemize}

\pagebreak
\section{Report structure}
This report is divided into 6 chapters, describing the above contributions and the context from which they arise:

\begin{description}
	\item[Chapter~\ref{ch:background}]
	A review of the fundamentals needed for time-tiling in general, including its motivation and consequences.
	A survey of work related to stencil compilers, approximate solvers of partial differential equations, in particular the abstractions that they make available to the user.

	\item[Chapter~\ref{ch:devito}]
	Details of the existing Devito compilation process which will be affected by time-tiling, and its existing implementation of spatial tiling.

	\item[Chapter~\ref{ch:implementation}]
	The implemented time-tiling transformation.
	Explanations of the design choices made, and modifications to the auto-tuner to ensure time-tiling is used effectively.
	Proposals for further implementation work arising from time-tiling.

	\item[Chapter~\ref{ch:evaluation}]
	An analysis of the effectiveness of the implemented transformation.
	Evidence of large performance improvements, and the conditions under which they are most and least significant.
	The aforementioned estimator for arithmetic intensity under time-tiling and its evaluation.
	Finally, areas for improvement in identifying and realising greater decreases in runtime.

	\item[Chapter~\ref{ch:conclusion}]
	The conclusion of the work, and a discussion of future work to be done alongside the streamlining of this optimisation process.
\end{description}

\paragraph{Time-tiling in Devito}
The implemented transformation from this work is currently being integrated into the Devito codebase; based on our evaluation, it is expected to bring significant performance improvement to customer applications.

	\documentclass[thesis.tex]{subfile}



\chapter{Background and related work}
\label{ch:background}

We first provide an overview of loop nest optimisations, techniques, and analyses that we have applied, with specific reference to time-tiling.
This forms the basis and context for the entire report, and specifically informs our overview of Devito (Section~\ref{sec:devito}) and the survey of related work, which composes the remainder of this chapter (Section~\ref{sec:survey}).

This project extends a well-established idea from compiler theory, \emph{tiling}, to another dimension (time) in Devito.
When tiling, particular attention need be paid to data dependences, locality, and parallelism of the resultant code.
Despite the confluence of factors, tiling is a remarkably simple concept.
Nevertheless, it is challenging to implement correctly in sufficient generality for Devito, being largely restricted to research compilers apart from OPS; in particular, we implement \emph{parameterised} time-tiling, which allows tile sizes to be changed at runtime.


\section{Terminology}
In addition the terms defined in the sections that follow, we use the following definitions widely in the context of this work:

\paragraph{Iteration space}
The \emph{iteration space} is the discretised problem domain that we iterate over.
In general, Devito handles hyperrectangular iteration spaces, but in this work it is sufficient to consider three or fewer spatial dimensions and a time dimension.
All remarks in this work can be considered to generalise to higher spatial dimensions unless otherwise stated.

\paragraph{Stencil}
A \emph{stencil} defines a computation on an iteration space.
In particular, note that a stencil will define the value of a point in terms of its spatial neighbours in previous time iterations.
It may be helpful to recall that we are employing finite-difference methods, and Taylor's theorem in higher dimensions.

\paragraph{Space order}
The \emph{space order of a stencil} defines the precision of the resulting computation.
A stencil with higher space order will use data from, hence have a dependence on, more distant spatial neighbours.
The \emph{radius} of a stencil refers to the distance of the furthest spatial dependency.

\paragraph{Arithmetic intensity}
The \emph{arithmetic intensity of a stencil} is the number of floating point operations performed per byte of data required, given in floating point operations per byte.

\section{Loop tiling}
\label{sec:bg-loop-tiling}

\subsection{Motivation}
The bulk of computation lies in loops~\cite{aho-cptt}.
Finite-difference methods are no exception, and in such computations, nested loops (`loop nests') are frequently encountered.
Loop nest optimisations seek to transform a loop, possibly changing its execution order to improve data locality, parallelism, or otherwise avoid unnecessary operations.

\subsection{Insight}
To exploit data locality, we must use data before it gets evicted from the cache; ideally, data is not loaded into the cache more than once.
Caching and predicting reuse is a complex process \cite{lam91}; empirically, reuse does occur within sufficiently small iteration spaces.
We therefore contrive small iteration spaces by partitioning the original space into smaller tiles (Figure~\ref{fig:tiled-space}).

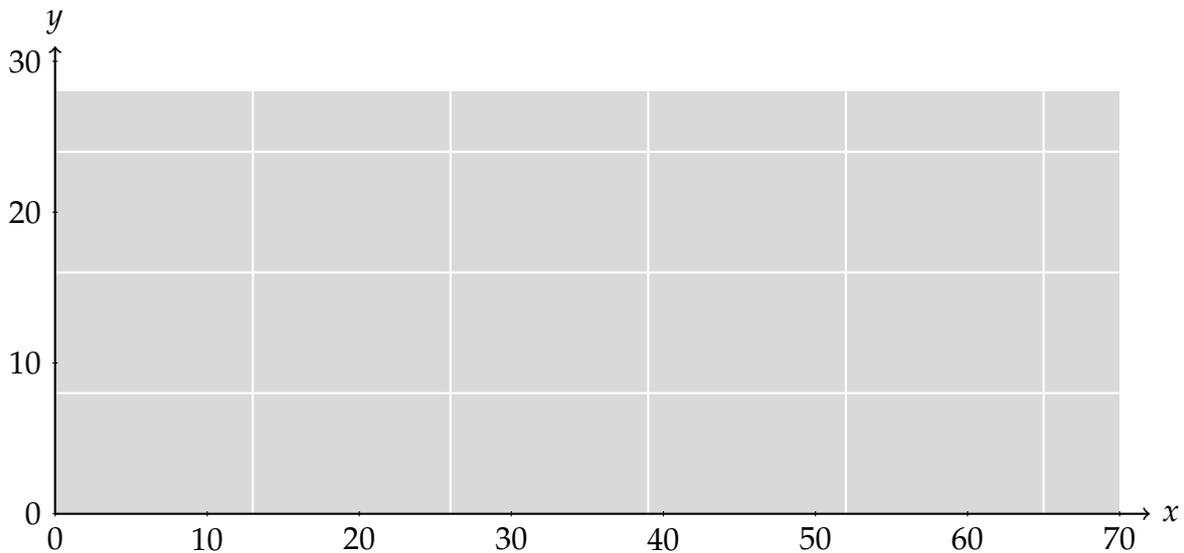
\begin{figure}[ht]
	\centering
	\begin{tikzpicture}
	\fill[gray!30!white] (0,0) rectangle (14,5.6);
	\draw[xstep=2.6cm,ystep=1.6cm,white,thick] (0,0) grid (14,5.6);

	\draw[thick,->] (0,0) -- (14.4,0) node[right]{$x$};
	\draw[thick,->] (0,0) -- (0,6.2) node[above]{$y$};
	\foreach \x in {0,10,20,30,40,50,60,70}
		\draw (\x*0.2 cm,1pt) -- (\x*0.2 cm,-1pt) node[anchor=north] {$\x$};
	\foreach \y in {0,10,20,30}
		\draw (1pt,\y*0.2 cm) -- (-1pt,\y*0.2 cm) node[anchor=east] {$\y$};
	\end{tikzpicture}
	\caption{Tiles over an iteration space. Note that the tile size need not be the same in each dimension, or divide the extent of the iteration cleanly.}
	\label{fig:tiled-space}
\end{figure}

Loop tiling is also commonly known as \emph{blocking}, or perhaps less transparently \emph{strip-mine and interchange}, as tiling is typically achieved through these two transformations.

\paragraph{Additional motivation}
As a remark, the stencils resulting from finite-difference methods tend to have high \emph{arithmetic} intensity, while tiling is used to reduce \emph{memory} pressure.\footnote{Thus stencils may not benefit from memory reuse if the bottleneck occurs at the CPU.}
However, as we note later, arithmetic intensity can be decreased (at the expense of increasing memory pressure) by eliminating common sub-expressions and other transformations, which would ordinarily result in redundant computation.
Further, tiling may also enable other transformations, such as loop-invariant code motion, which again reduces redundant computation.

\subsection{Strip-mining}
Named after the mining practice, strip-mining involves dividing a dimension of the iteration space into strips (Figure~\ref{lst:stripmine-basic}).\footnote{However, you cannot divide a dimension into lateral strips, only sequential ones.}
By itself, strip-mining does not change the execution order; it is a gateway to further transformations.

\begin{figure}[ht]
\begin{lstlisting}
for (int x = x_start; x < x_end; x++) {
  A[x] = B[x-1] + B[x+1];
}

for (int x_blk = x_start ; x_blk < x_end; x_blk += x_blk_size) {
  for (int x = x_blk; x < min (x_end, x_blk + x_blk_size); x++) {
    A[x] = B[x-1] + B[x+1]; // loop body unchanged
  }
}
\end{lstlisting}
	\caption{A regular loop, and the same loop strip-mined over the variable \texttt{x}. \texttt{x\_start} and \texttt{x\_end} are chosen to prevent out-of-bounds accesses. In case the tile (block) size does not evenly divide the extent of the iteration, the \texttt{min} function avoids the need for remainder loops (discussed in Section~\ref{sec:remainder-loops}). We will abbreviate the variable names in subsequent examples.}
	\label{lst:stripmine-basic}
\end{figure}

\begin{framed}
	Throughout this work, we will refer to the outer loop (variable \texttt{x\_blk} in Figure~\ref{lst:stripmine}) as the \emph{tile} loop, and the inner loop (variable \texttt{x} in the same figure) as the \emph{incremental} loop.
\end{framed}

We will need more loops to perform an interchange.
Figure~\ref{lst:stripmine} illustrates a loop that has been strip-mined in two dimensions.

\begin{figure}[ht]
	\begin{lstlisting}
for (int x_blk = x_s; x_blk < x_e; x_blk += x_bs) {
  for (int x = x_blk; x < min(x_e, x_blk + x_bs); x++) {
    for (int y_blk = y_s; y_blk < y_e; y_blk += y_bs) {
      for (int y = y_blk; y < min(y_e, y_blk + y_bs); y++) {
        A[x][y] = B[x][y] + B[x][y+1];
      }
    }
  }
}
	\end{lstlisting}
	\caption{Strip-mining a loop nest iterating over variables \texttt{x} and \texttt{y}. Offsets have been omitted here.}
	\label{lst:stripmine}
\end{figure}

\subsection{Loop interchange}
\label{sec:interchange}
Loop interchange is based on the observation that a change in execution order does not change the correctness of a strip-mined program.
We will change the order of the loops to iterate over the tiles, then within them (Figure~\ref{lst:interchange}).

\begin{figure}[ht]
\begin{lstlisting}
for (int x_blk = x_s; x_blk < x_e; x_blk += x_bs) {
  for (int y_blk = y_s; y_blk < y_e; y_blk += y_bs) {
    for (int x = x_blk; x < min(x_e, x_blk + x_bs); x++) {
      for (int y = y_blk; y < min(y_e, y_blk + y_bs); y++) {
        A[x][y] = B[x][y] + B[x][y+1];
      }
    }
  }
}
\end{lstlisting}
	\caption{The loop nest of Figure~\ref{lst:stripmine}, with the \texttt{x} and \texttt{y\_blk} loops interchanged. This is a \emph{tiled} loop.}
	\label{lst:interchange}
\end{figure}

This is valid when each point in the iteration space does not depend on the values calculated in the same iteration.
Therefore, one must be extremely careful that no data dependences cross boundaries between tiles; if they do, they must be permitted to cross only in one direction, and the tiles must be scheduled in that order.
This will become clear in Section~\ref{sec:bg-skewing}.

\section{Tiling in the time dimension}
\label{sec:bg-time-tiling}

\subsection{Motivation}
Many problems involving finite difference methods are computationally bounded, rather than bounded by memory throughput.
However, it is possible to reduce the operation count by exploiting the structure of expressions computed at the cost of increased memory pressure~\cite{fabio-memory}.

We are therefore interested to perform time-tiling in addition to the existing spatial tiling to alleviate the increased memory pressure and realise significant performance gains, possibly 27.5\% in Devito alone~\cite{dylan}.
This improvement is significant over the optimisation from tiling in all dimensions apart from time~\cite{pluto}.

In spatial tiling, we exploited data locality within the computation of a single time iteration.
In contrast, time-tiling exploits data locality between time iterations.

\subsection{Skewing}
\label{sec:bg-skewing}
In Section~\ref{sec:interchange} we stated that interchange is valid when data dependences do not cross boundaries between tiles.
This is clear, as if there are no inter-tile dependences, the tiles can be executed in any order.

\begin{framed}
	A \emph{data dependence} occurs when two statements reference a datum, and at least one statement changes it.
	Failure to preserve the execution order of the statements is known as \emph{dependency violation}.
\end{framed}

To preserve the dependence, we must preserve the order in which these statements are executed.
Figure~\ref{fig:dependence} illustrates how dependences may look in a (1-dimensional) iteration space similar to the problems we discuss later.
Dependency violation occurs with an invalid ordering which overwrites data yet to be read, or requires data that has not been calculated yet.

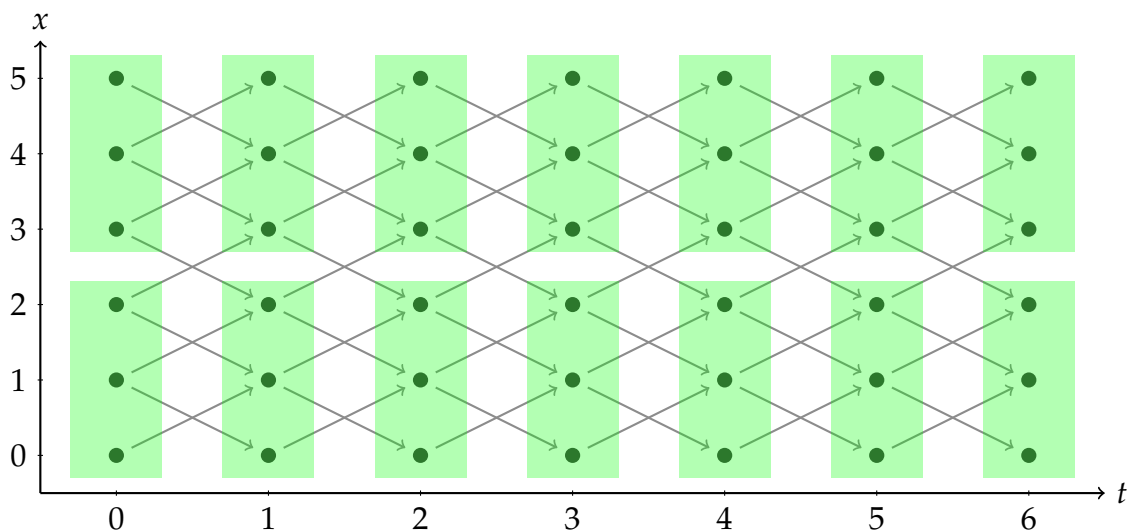
\begin{figure}[ht]
	\centering
	\begin{tikzpicture}
	\draw[thick,->] (-1,-.5) -- (13,-.5) node[right]{$t$};
	\draw[thick,->] (-1,-.5) -- (-1,5.5) node[above]{$x$};

	\foreach \t in {0,1,2,3,4,5,6}
	\foreach \x in {0,1,2,3,4,5}
		\fill[darkgray] (\t*2,\x) circle (0.1);

	\foreach \t in {0,1,2,3,4,5}
	\foreach \x in {1,2,3,4,5}
		\draw[thick,->,darkgray!60] (\t*2+.2,\x-.1) -- (\t*2+1.8,\x-.9);

	\foreach \t in {0,1,2,3,4,5}
	\foreach \x in {0,1,2,3,4}
		\draw[thick,->,darkgray!60] (\t*2+.2,\x+.1) -- (\t*2+1.8,\x+.9);

	\foreach \t in {0,1,2,3,4,5,6}
	\foreach \x in {0,3}
		\fill[green,opacity=0.3] (\t*2-.6,\x-.3) rectangle (\t*2+.6,\x+2.3);

	\foreach \t in {0,1,2,3,4,5,6}
		\draw (\t*2,1pt-.5cm) -- (\t*2,-1pt-.5cm) node[anchor=north] {$\t$};
	\foreach \x in {0,1,2,3,4,5}
		\draw (1pt-1cm,\x) -- (-1pt-1cm,\x) node[anchor=east] {$\x$};
	\end{tikzpicture}
	\caption{An iteration space with data dependences indicated by a forward arrow for a value derived from a dependence. It would not be valid to interchange loops over the \texttt{t} and \texttt{x} dimensions here.}
	\label{fig:dependence}
\end{figure}

We employ skewing to make the interchange valid (Figure~\ref{fig:dependence-skew}).
This solves the dependency problem~\cite{boulet98}, and allows for loop interchange.

\begin{figure}[!ht]
	\centering
	\begin{tikzpicture}
	\draw[thick,->] (-1,-.5) -- (13,-.5) node[right]{$t$};
	\draw[thick,->] (-1,-.5) -- (-1,5.8) node[above]{$x$};

	\foreach \t in {0,1,2,3,4,5,6}
	\foreach \x in {0,1,2,3,4,5}
		\fill[darkgray] (\t*2,\x/2+\t/2) circle (0.1);

	\foreach \t in {0,1,2,3,4,5}
	\foreach \x in {1,2,3,4,5}
		\draw[thick,->,darkgray!60] (\t*2+.2,\x/2+\t/2) -- (\t*2+1.8,\x/2+\t/2);

	\foreach \t in {0,1,2,3,4,5}
	\foreach \x in {0,1,2,3,4}
		\draw[thick,->,darkgray!60] (\t*2+.2,\x/2+\t/2+.05) -- (\t*2+1.8,\x/2+\t/2+.9);

	\foreach \t in {0,1,2,3,4,5,6}
	\foreach \x in {0,3,6,9}
		\fill[green,opacity=0.3] (\t*2-.6,\x/2-.15) rectangle (\t*2+.6,\x/2+1.15);

	\foreach \t in {0,1,2,3,4,5,6}
		\draw (\t*2,1pt-.5cm) -- (\t*2,-1pt-.5cm) node[anchor=north] {$\t$};
	\foreach \x in {0,1,2,3,4,5,6,7,8,9,10,11}
		\draw (1pt-1cm,\x/2) -- (-1pt-1cm,\x/2) node[anchor=east] {$\x$};
	\end{tikzpicture}
	\caption{The same iteration space skewed by a factor of \texttt{t} in the \texttt{x} dimension. Note that the tiling is now valid (as long as we avoid out-of-bounds accesses) and we can execute the tiles in either dimension first, and that we can merge tiles in the \texttt{t} dimension.}
	\label{fig:dependence-skew}
\end{figure}
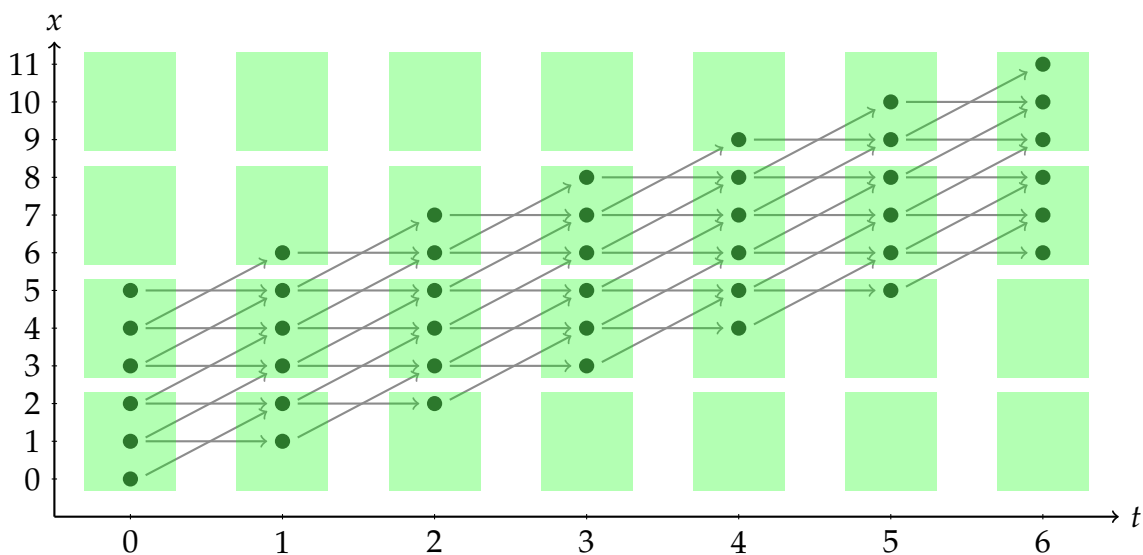

\subsubsection{Validity of skewing factors}

We have seen that skewing is a tool to enable loop interchange, and that interchange is legal when dependences only cross tile boundaries in one direction.
To ensure this, a sufficiently large skewing factor must be used.
We claim that the maximum spatial dependence distance is a valid skewing factor~\cite{xue97-tiling}, i.e.~in a stencil \texttt{A[t][x] = c0*A[t-1][x] + c1*A[t-1][x-1] + ... + cn*A[t-1][x-n];}, the minimum valid skewing factor is \texttt{n}.

As a remark, in this work we have largely decided to skew all spatial dimensions with this skewing factor, due to its simplicity, and the stencils under consideration.
However, it is equally valid to skew each spatial dimension by the dependence distance \emph{for that particular dimension}~\cite{schreiber-dongarra}.

\section{Devito}
\label{sec:devito}

Devito~\cite{devito} is a domain-specific language and code generation framework for solving PDEs with finite-difference methods~\cite{devito-web}.
Its main use case is building solvers for differential equations from high-level mathematical expressions, written using the symbolic library \emph{SymPy}, and is targeted at the domain of seismic imaging.

\subsection{Motivation}
Efficient computation demands that we use every optimisation at our disposal.
As computer architecture changes, introducing new instructions, GPUs and distributed systems, specialist research compilers implementing subsets of related optimisations do not suffice for high-performance applications.

An equal part of optimisation is reducing the turnaround time in modelling.
Providing an interface that domain specialists are familiar with reduces friction, and removes the need to construct stencils by hand.
This also reduces errors by simplifying the inputs that they provide.

This combination of ease of use through the Python and \emph{SymPy} interface and the generation of optimised and fast C code without user intervention make Devito an attractive tool.

\subsection{Domain-specificity}
Polyhedral compilers such as PLUTO~\cite{pluto}, with its backend CLooG~\cite{cloog-isl} are able to apply tiling over generic loop nests.
Further, the iteration spaces which they handle are unions of convex polyhedra, again more general than the use cases Devito is likely to encounter.
This generality is not necessary when merely applying finite-difference methods to solve differential equations, not least when targeting a specific domain which uses such methods.

We can instead make use of this specificity to make informed choices in our optimisations, or streamline an auto-tuning routine.
Further, we are able to combine different optimisations, at the stencil level and progressively lower levels.
Finally, considering likely use cases, it may be desirable to target compilation for architectures not supported by more general compilers, such as GPUs.

\subsection{Layers of abstraction}
The separation of symbolic expressions and the underlying C code to which they are compiled is key to Devito's comprehensibility and usage.
By progressive manipulation, optimisation and modification is possible at many layers.
A further benefit as a more general software engineering practice is the possibility of small unit tests for individual components.

This also allows the integration of other tools, which may fit a user's needs or architectures better, and enables easy benchmarking of Devito's performance against said tools.
For instance, YASK (Section~\ref{sec:yask}) is being integrated as a compilation backend.

An important feature for seismic imaging is sparse-point interpolation.
Devito handles this by exposing powerful lower-level APIs to allow the construction of comprehensive representations which would be tedious or impossible at higher levels of abstraction.

\subsection{Architecture}
The Devito compilation process (Chapter~\ref{ch:devito}) can be divided into several stages:

\begin{enumerate}
	\item Construction of stencil equations (symbolic kernel)
	\item Grouping of expressions
	\item CSE and indexing of stencil (the Devito symbolic engine (DSE), Section~\ref{sec:dse})
	\item Loop optimisation and other transformations (the Devito loop engine (DLE), Section~\ref{sec:dle})
	\item Code generation
\end{enumerate}

Our time-tiling transformation will occur in the DSE and DLE, covered in detail in the respective sections.

\section{Survey of related work}
\label{sec:survey}

The following sections provide an overview of tools related to Devito and of interest to our investigation.
In particular, they identify instances of time-tiling (or equivalent transformations) which have been solved and the insights required, and analyse their applicability to time-tiling in Devito.

\subsection{Halide}
Halide was conceived as a representation for image processing pipelines.
Many image processing algorithms are similar to stencils: Halide specifically deals with overlapping stencils; this overlap can be compared to iteration in the time dimension.

\subsubsection{Insight}
It is possible to separate image processing algorithms (``filters'') and their schedules~\cite{halide12}.
Optimisation for an architecture then becomes an exercise in optimising the schedule and not the filters, which are reduced to kernels.\footnote{See {https://github.com/halide/CVPR2015/blob/master/blur.cpp} for an example}
Halide is therefore a stencil compiler.

Modifying schedules is analogous to our loop optimisations: filters can be vectorised, tiled, interchanges are possible, etc.
Halide further provides an auto-tuner which estimates, among other things, arithmetic intensity of filters and loop transformations which Devito also employs~\cite{halide13,halide-sched}.
Part of its analysis examines the trade-off between additional computation and memory traffic.

\subsubsection{Applicability}
The problem domain, image processing, that Halide is concerned with may appear very different to that of differential equations.
However, the underlying natures of the computations are very similar: use of stencil kernels, trade-off between redundant computation and memory pressure, etc.
Analogously to time-tiling, Halide is able to compute tiles across filters.

Likewise, in Devito, \emph{each time iteration} may contain several stencil kernels applied consecutively.
To extend the analogy, Halide is concerned with tiling within \emph{a single} timestep, while this project investigates tiling over \emph{multiple} timesteps.

Domain-specific frameworks bring insight to a problem which general-purpose compilers may not possess; the separation of algorithm and schedule, while touted as novel, is essentially what every compiler applies during transformations such as loop interchange.

\subsection{Pochoir}
\label{sec:pochoir}
Pochoir is a compiler for stencil computations focussed on utilising parallelism and multithreading.
Stencils are defined with provided C++ templates, from which the computations are generated.
Pochoir implements a two-phase compilation strategy: the first involves compilation with the Pochoir template library, which ensures compliance and compatibility with the library. At this stage, one may debug the non-optimised code.
The second is the optimisation phase using the Pochoir compiler\footnote{The Pochoir compiler will also invoke the user's Cilk Plus compiler.}~\cite{pochoir}.

\subsubsection{Insight}
\paragraph{Cache-oblivious algorithms}
These seek to eliminate tuning of cuts (tile size) based on cache properties including cache sizes or replacement policies, reasoning that the correct cache-oblivious algorithm will achieve (asymptotically) the same performance~\cite{frigo-cacheoblivious}.

Pochoir uses a cache-oblivious algorithm based on parallel cuts.
These \emph{hyperspace} and \emph{time cuts} decompose the iteration spaces (``zoids'') into smaller spaces recursively;
they are chosen to improve parallelism while maintaining cache efficiency~\cite{frigo-cacheoblivious-alg}.

Thanks to cut dependence analysis, the resultant trapezoids can span iterations of a time loop. Note that the domain which Pochoir operates on are unions of convex polyhedra, which are more general than the hyperrectangular iteration spaces described so far.

\subsubsection{Applicability}
The Cilk Plus framework, which Pochoir uses, is intended to produce optimal scheduling for parallel tasks, and cache-oblivious algorithms may be optimal up to a constant factor, which would be interesting when targeting multiple platforms.
Nevertheless, significant speedup can occur with \emph{cache-aware} algorithms, which employ tuning for multiple levels of cache~\cite{kowarschik-cacheaware}.
Indeed, due to a difficulty in choosing suitable base case sizes for the interior trapezoids, Pochoir includes an auto-tuner for this purpose.
This is especially significant to domain specialists who are experimenting and changing models, or those who are performing stencil computations over large iteration spaces.

Finally, the Cilk Plus framework is in the process of deprecation~\cite{cilkplus-migrate-web}, and Pochoir is not presently maintained.
While the concepts behind it may be sound and applicable in specific circumstances, it would be unsuitable for integration with Devito.

\subsection{YASK}
\label{sec:yask}

\emph{Yet Another Stencil Kernel} is a framework from Intel Corp. that transforms and optimises stencil kernels, especially targeting the Xeon Phi platform~\cite{yask-web}.
Like Pochoir, YASK provides C++ templates for stencils.
Similarly to Halide, it uses a genetic-algorithm based search for its auto-tuner.

It is intended to function both as a platform for domain specialists to experiment with (higher-level) optimisations, as well as for the compilation of high-performance kernels.

\subsubsection{Insight}
Stencil kernels rapidly grow complicated.
Once optimisations have been applied (possibly through a specialised stencil compiler) to stencils, general-purpose compilers are less able to explore the trade-offs of further optimisation which they would ordinarily apply, such as parallelism.
YASK aims to solve this by, among other things, implementing a two-stage compiler (stencils and loops) and tuning for the execution environment~\cite{yask-paper}.
We draw a comparison to Devito here.

In contrast to polyhedral research compilers such as PLUTO~\cite{pluto}, YASK has a specific focus on combining optimisations such as vector folding~\cite{yount-vector} with the loop optimisations which they perform.

\subsubsection{Applicability}
YASK was originally created to implement vector folding, a technique which simultaneously takes advantage of SIMD instructions and reduces memory bandwidth usage.
This is especially relevant in Devito, which has the transformation of arithmetically intensive computations into memory intensive computations as a principle.

Further, with its emphasis on integrating different optimisations, rather than investigating them separately, YASK is a practical tool for a workflow involving stencils.
There would be no need to apply a separate stencil compiler, then a polyhedral compiler, without the guarantee that the latter would understand and be able to build open the complexities introduced by the former.
One might remark, however, that the same applies when constructing stencils from differential equations: instead they are reduced to a form (C++ templates) that the compiler understands.

As noted in Section~\ref{sec:devito}, Devito is in the process of integrating YASK as a compilation backend.

\subsection{OPS}
The \emph{Oxford Parallel library for Structured mesh solvers}~\cite{ops-main} is a domain specific language that, like Pochoir and YASK, provides an abstraction and library for optimisation and compilation of stencils, although the stencils can be defined in C and Fortran as well as C++.
OPS operates on \emph{multi-block structured meshes}, which are a union of hyperrectangular blocks on a grid, with a focus on parallelism.

Like Devito, it is intended to enable easy code maintenance, except at the stencil level, rather than from the models and equations; one use case is the migration of existing stencils for use with OPS.

\subsubsection{Insight}
Apart from the tenets of maintainability and abstraction of optimisation which are familiar to the reader, OPS is designed to compile to different target platforms, and implements a novel lazy evaluation scheme.

In lazy evaluation, data is only evaluated when it is required; OPS uses the approach in distributing computation to parallel nodes such as GPUs~\cite{ops-gpu}, which have high memory bandwidth.
The intention is to optimise the scheduling of computation across nodes.

\subsubsection{Applicability}
The ability to target different platforms is a priority in Devito, and OPS highlights that understanding architectural differences is very important in developing optimising compilers.
While the runtime overheads of lazy evaluation are well-known from functional programming languages, OPS attempts to use the concept to optimise scheduling the execution of tiles for distributed memory systems.

Additionally, OPS has been tested against real-world problems in fluid dynamics, presenting a time-tiling algorithm in June 2017, albeit noting that it still falls short of hand-optimised code and polyhedral compilers~\cite{ops-tiling}.
One may note that hand-optimisation is well-established and very effective in computational fluid dynamics, and more importantly the comparison platform was not a distributed memory system which OPS targets, hence this may not be an issue.

As with Devito, a time iteration may contain multiple stencil kernels to be applied sequentially.
The main differentiating factors are the use of lazy evaluation for scheduling and consequential focus on distributed memory platforms and that it operates on stencils, which need to be generated from PDEs.

Like OPS, Devito is developed with a focus on real-world applications, also requiring rigorous analysis, with more complicated stencils than traditionally used to benchmark polyhedral compilers such as PLUTO.
In this work, we evaluate our implementation of time-tiling against a family of stencils generated by the acoustic wave equation.

\subsection{Firedrake}
Firedrake is a tool for solving partial differential equations on unstructured meshes using the \emph{finite-element method}~\cite{firedrake}.
It is aimed at simulation and modelling of systems through PDEs, notably separating optimisations into different layers of abstraction.

FEniCS~\cite{fenics}, another tool in the domain of finite-element solvers, separates the usage of finite-element methods and their implementation; Firedrake takes this a step further while retaining the domain-specific language and focus which it derives.

\subsubsection{Insight}
Successfully optimising the solution of PDEs involves expertise at many levels, from numerical analysis of the equations and mesh generation, to the loop optimisations discussed here.
Just as we separate stencil optimisations from polyhedral ones, we should also abstract these layers.

Again, this makes the \emph{optimised} solution of partial differential equations accessible to domain specialists without particular expertise in computer architecture, transformation of expressions into efficient kernels~\cite{coffee}, or mesh generation.

\subsubsection{Applicability}
Through its many layers of abstraction, Firedrake provides a natural way to access lower-level structures such as kernels and the underlying data structures.
Like Devito, this allows specialists to manipulate the computation not just at the highest level of the model, but also at levels which the specialist may have insight to, but would not be otherwise exposed.

While many of the problems which Firedrake deals with are not directly related to Devito (such as unstructured meshes), the concept of abstraction is sound and very relevant to software engineering.
Of particular relevance is the granularity to which abstractions are helpful at the lower levels in which we are interested, the use of kernels (albeit not stencils), and the discretisation of operators.

Hence Firedrake comes far closer to Devito than the previously-discussed tools, as it covers the full process of computation, from the models and differential equations rather than the resultant kernels.
Although the main optimisation that we discuss, time tiling, functions at the kernel level, Firedrake gives a broader view of the context of the problem to be solved, and the model in which Devito resides.

\section{Summary}
In this chapter, we have given an overview of the core optimisation concepts explored and terminology used throughout this work.
Further, we discussed the motivation and context of Devito specifically, its overall usage and architecture (elaborated in Chapter~\ref{ch:devito}), and the importance of the time-tiling transformation; these will inform our evaluation (Chapter~\ref{ch:evaluation}).
Finally, we have examined related work in the form of tools employing transformations relevant to time-tiling and stencil computations, and most importantly, the insights which they bring to our implementation (Chapter~\ref{ch:implementation}).

	\documentclass[thesis.tex]{subfile}

\chapter{Space-tiling and the Devito compilation process}
\label{ch:devito}

We reviewed Devito's motivation and overall architecture in Section~\ref{sec:devito}.
In order to inform our implementation of time-tiling, this chapter examines the details of the existing Devito compilation process pertinent to our implementation of time-tiling.
We examine the transformations that are relevant to the time-tiling procedure, specifically those that occur immediately before and after tiling and skewing.
In particular, we review the existing implementation of spatial tiling (Section~\ref{sec:spatial-tiling}).

In general, we have assumed that Devito only performs valid transformations, which we verify against its extensive test suite.
Therefore, we are only concerned with the transformations that occur between skewing and tiling.

\section{Overview}
Devito generates computations using finite-difference methods to solve differential equations, which are defined using \emph{SymPy}.
These are defined as operators, which define a problem domain and the equations to be applied (Figure~\ref{lst:devito-op}).
A stencil is then derived, optimised, and placed into loops, which are further optimised.

\begin{figure}[!ht]
\begin{lstlisting}[language=Python]
def operator(shape, **kwargs):
    grid = Grid(shape=shape)
    spacing, a, c = 0.1, 0,5, 0,5
    dx2, dy2 = spacing**2, spacing**2
    dt = dx2 * dy2 / (2 * a * (dx2 + dy2))

    # Allocate the grid and set initial condition
    u = TimeFunction(name='u', grid=grid, time_order=2, space_order=2)
    u.data[0, :] = np.linspace(1e-20, 2e-20,
                               num=reduce(mul, shape)).reshape(shape)

    # Derive the stencil according to devito conventions
    eqn = Eq(u.dt, a * (u.dx2 + u.dy2) - c * (u.dxl + u.dyl))
    stencil = solve(eqn, u.forward, rational=False)[0]
    op = Operator(Eq(u.forward, stencil), **kwargs)

    # Execute the generated Devito stencil operator
    op.apply(u=u, t=10, dt=dt)
    return u.data[1, :], op
\end{lstlisting}
	\caption{Defining and applying an operator in Devito. Some set-up is required to define the equation, here a Laplace operator in the \(x,y\) dimensions. In actual usage, the grid (\texttt{u.data[0]}) would be the input domain. The stencil is derived, then applied to the grid.}
	\label{lst:devito-op}
\end{figure}

Our starting point is the stencil that Devito has generated, following its transformation from stencil optimisation in the DSE until the point where loop tiling is applied in the DLE.
We have chosen to restrict the scope of this chapter to these components, which are pertinent to the time-tiling transformation that we implement in Chapter~\ref{ch:implementation}.

\section{The Devito symbolic engine (DSE)}
\label{sec:dse}

The Devito symbolic engine is responsible for stencil optimisation and common sub-expression elimination (CSE).
At this point the indices and loops have not been generated yet, but we have generated the stencil and know its data dependencies.

Although we are working with loop transformations, we will perform skewing (Section~\ref{sec:skewing}) in the DSE, as skewing is a modification to the canonical indexing scheme.

\subsection{Common sub-expression elimination (CSE)}
The purpose of CSE is to reduce redundant computation, by removing duplicate expressions~\cite{muchnick-acdi}.
Figure~\ref{lst:cse-valid} illustrates a situation in which CSE would be used.
In this situation, CSE would remove the two expressions, replacing it with one that subtracted \texttt{0.5F*u[time][x][y][z]}.

\begin{figure}[!ht]
\begin{lstlisting}
float tcse0 = -1.0F*u[time][x][y][z] + ...
              + .5F*u[time][x][y][z];
\end{lstlisting}
	\caption{Code (from Devito) eligible for common sub-expression elimination. Formatting is my own.}
	\label{lst:cse-valid}
\end{figure}

In the DSE, this is also performed at a higher level.
There is a balance to be made here: time-invariant expressions are first separated and assigned to temporaries, avoiding needless CSE, thus decreasing code-generation time.

\subsection{Indexing}
Indexing in Devito is the transformation of a stencil equation with distance co-ordinates to an expression of an indexed array accesses.
It is a transformation between internal representations of the stencil.

Despite being considered a loop transformation, we have chosen to perform skewing before indexing occurs, for reasons which will be clear in Section~\ref{sec:impl-cse}.
However, it is equally valid and perhaps more intuitive to skew after indexing.

\section{The Devito loop engine (DLE)}
\label{sec:dle}

The DLE performs loop transformations including loop fission and tiling, while also marking loops to be executed in parallel or that denormal numbers should be flushed.
Before it is invoked, the loops are built from the previously-manipulated expressions; clearly it would not be possible to implement tiling before this stage.

\subsection{Loop fission}
Loop fission splits a loop to improve data locality.
If the body of a loop nest contains code that can be split into otherwise independent loops, it may be a candidate for loop fission (Figure~\ref{lst:fission-valid}), especially if the body can be computed in parallel~\cite{page2009}.
By evaluating the loops one after another, the amount of cache available to each computation increases, potentially resulting in a speedup.

\begin{figure}[!ht]
\begin{lstlisting}
for (int x = x_s; x < x_e; x++){
  A[x] = A[x-1] + A[x-2];
  B[x] = B[x-1] + B[x-3];
}
\end{lstlisting}
	\caption{A loop nest on which loop fission would be valid.}
	\label{lst:fission-valid}
\end{figure}

We can easily see that this does not affect spatial tiling in Devito, as it has no impact on array indices, since the resultant loops appear in the same nest as the original loop, with the variables duplicated.
Likewise, this holds equally for time-tiling.

\subsection{OpenMP pragmas and parallelisation}
Another particularly relevant operation of the DLE is the addition of OpenMP pragmas indicating that a particular loop may be parallelised~\cite{openmp-spec}.

While this transformation is performed after tiling, it is pertinent as loops must be marked as parallelisable in the intermediate representation.
Loop tiling generates new loops from old, resulting in loop nests that are twice as deep.
It is important to note which of the original loops can be parallelised, and consequently, which of the tiled loops can be parallelised.
This is not a major concern for spatial tiling, as seen in Section~\ref{sec:spatial-tiling}, but will later be important in time-tiling.

\section{Spatial tiling in Devito}
\label{sec:spatial-tiling}

This section documents the existing spatial tiling transformation in Devito, and outlines some important differences and concepts not previously discussed.
Note that in the current transformation, only spatial dimensions may be tiled, and the innermost dimension may be vectorised instead.

With spatial tiling, dependencies do not cross tile boundaries, instead only referencing values computed in the previous time iteration.
Therefore, skewing is not required.

\subsection{Parameterisation}
A challenge when tiling general loops as required in Devito parameterisation of loop bounds, and the correct propagation of these bounds.
As seen in the examples of Section~\ref{sec:bg-loop-tiling}, the tile size for each dimension is a variable such as \texttt{x\_blk\_size}.
This is a \emph{parameterised} tile, meaning that tile sizes can be changed at runtime, which enables features such as auto-tuning, which is discussed later.
Therefore, there is no need to re-optimise and re-compile a stencil from a change in tile size; resulting in considerable time savings.

\subsection{Auto-tuner}
While not strictly part of the compilation process, the Devito auto-tuner is relevant as it uses the above-mentioned tile size parameters to experimentally determine an optimal combination of tile sizes for a particular stencil and domain.

In particular, an implementation of time-tiling would have to expose similar parameters, supplemented with an additional parameter for the time dimension.

\subsection{Determination of tile sizes}
One notes that the loop tiling method includes code to validate user-provided tile sizes, or provide default otherwise.
Nevertheless, this is not relevant to the discussion of time-tiling in this report, apart from the straightforward addition of the time dimension.
Therefore, it is omitted from this section.

\subsection{Remainder loops}
\label{sec:remainder-loops}

When discussing loop tiling in Section~\ref{sec:bg-loop-tiling}, we constrained our tiles with \texttt{min} constraints.
To deal with the case when the tile size does not divide the extent of the iteration space, Devito instead implements \emph{remainder loops}, which Figures~\ref{fig:tiled-remainder-space} and~\ref{lst:remainder} illustrate.

\begin{figure}[!ht]
	\centering
	\begin{tikzpicture}
	\fill[gray!30!white] (0,0) rectangle (14,5.6);
	\draw[step=1.6cm,white,thick] (0,0) grid (14,5.6);
	
	\draw[thick,->] (0,0) -- (14.4,0) node[right]{$x$};
	\draw[thick,->] (0,0) -- (0,6.2) node[above]{$y$};
	
	\foreach \x in {0,10,20,30,40,50,60,70}
	\draw (\x*0.2 cm,1pt) -- (\x*0.2 cm,-1pt) node[anchor=north] {$\x$};
	\foreach \y in {0,10,20,30}
	\draw (1pt,\y*0.2 cm) -- (-1pt,\y*0.2 cm) node[anchor=east] {$\y$};
	
	\draw[xstep=12.8,ystep=4.8,black,thick] (0,0) grid (14,5.6);
	\draw (6,2.8) circle [radius=0.3] node {$1$};
	\draw (13.4,2.8) circle [radius=0.3] node {$2$};
	\draw (6,5.2) circle [radius=0.3] node {$3$};
	\draw (13.4,5.2) circle [radius=0.3] node {$4$};
	\end{tikzpicture}
	\caption{Tiles over an iteration space, with loop nests numbered 1--4. 1 indicates the main loop nest, the rest are remainder loop nests.}
	\label{fig:tiled-remainder-space}
\end{figure}
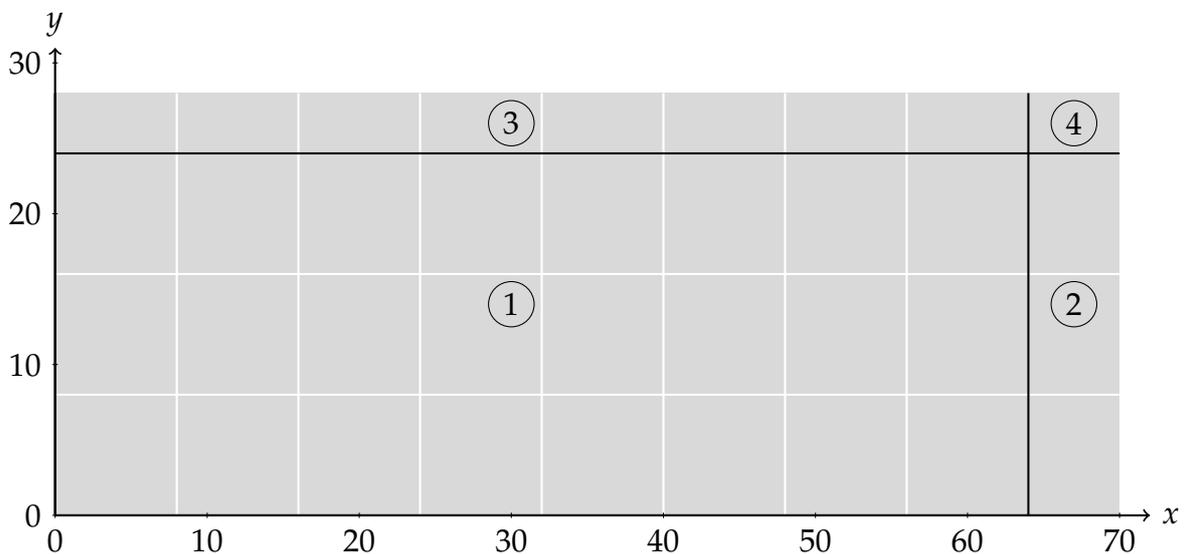

\begin{figure}[!ht]
\begin{lstlisting}
for (int x_blk = x_s; x_blk < x_e - (x_e-x_s)%x_bs; x_blk += x_bs)
  for (int y_blk = y_s; y_blk < y_e - (y_e-y_s)%y_bs; y_blk += y_bs)
    for (int x = x_blk; x < x_blk + x_bs; x++)
      for (int y = y_blk; y < y_blk + y_bs; y++)
        A[x][y] = B[x][y] + B[x][y+1];  // Nest 1

for (int x = x_e - (x_e-x_s)%x_bs; x < x_e; x++)
  for (int y_blk = y_s; y_blk < y_e - (y_e-y_s)%y_bs; y_blk += y_bs)
    A[x][y] = B[x][y] + B[x][y+1];  // Nest 2

for (int x_blk = x_s; x_blk < x_e - (x_e-x_s)%x_bs; x_blk += x_bs)
  for (int y = y_e - (y_e-y_s)%y_bs; y < y_e; y++)
    A[x][y] = B[x][y] + B[x][y+1];  // Nest 3

for (int x = x_e - (x_e-x_s)%x_bs; x < x_e; x++)
  for (int y = y_e - (y_e-y_s)%y_bs; y < y_e; y++)
    A[x][y] = B[x][y] + B[x][y+1];  // Nest 4
\end{lstlisting}
	\caption{Replacement of \texttt{min} constraints of Figure~\ref{lst:interchange} with remainder loops. First the main tiles, then the remainder in \texttt{x} then \texttt{y} dimensions, and finally the remainders in both dimensions. Nest numbering according to Figure~\ref{fig:tiled-remainder-space}. Braces removed for concision.}
	\label{lst:remainder}
\end{figure}

\section{Summary}
This chapter outlined the existing compilation process in Devito, highlighting details that require our attention in Chapter~\ref{ch:implementation}, the implementation of time-tiling.
These transformations and representations bely the abstractions that Devito uses to expose lower-level APIs, and they must be preserved fully when constructing a new transformation such as time-tiling.

	\documentclass[thesis.tex]{subfile}

\chapter{Implementation}
\label{ch:implementation}

Having previously discussed the goal of time-tiling, i.e.~to improve cache reuse by exploiting data locality between time iterations, this chapter presents our implementation of time-tiling in Devito, with reference to the design choices made.
For the most part, the implementation follows naturally from the previous chapters, although great care was required for the details, particularly the correct propagation of loop bounds under the transformations.

As outlined in Section~\ref{sec:bg-time-tiling}, we implement time-tiling in two stages: skewing in the DSE (Section~\ref{sec:skewing}), and tiling in the DLE (Section~\ref{sec:time-tiling}).
In addition to this, we have supplemented Devito test suite (Section~\ref{sec:impl-tests}) and auto-tuner (Section~\ref{sec:autotune}) for time-tiling.
Finally, we discuss outstanding implementation work for time-tiling in Devito (Section~\ref{sec:impl-further}).

\section{Skewing}
\label{sec:skewing}
A loop nest comprises several nested loops, each with its own loop variable.
We may assume the outermost loop holds the time variable without loss of generality.\footnote{Otherwise, ignore any outer loops.}
In our implementation, incremental loop variables are increased by a factor of time, and the index accesses of that variable are decreased by the same factor in the loop body (Figures~\ref{lst:skewing-simple} and~\ref{fig:skewed-loops-skewed-space}).
In this way, all accesses refer to the same data as before the transformation, ensuring its validity.

\begin{figure}[!ht]
\begin{lstlisting}
for (int t = t_s; t < t_e; t++)
  for (int x = x_s; x < x_e; x++)
    A[t][x] = A[t-1][x-1] + A[t-1][x+1];

for (int t = t_s; t < t_e; t++)
  for (int x = x_s + t; x < x_e + t; x++)
    A[t][(x-t)] = A[t-1][(x-t)-1] + A[t-1][(x-t)+1];
\end{lstlisting}
	\caption{The code transformation of skewing: an unskewed loop, and a skewed version of the same loop. Note that all index accesses still refer to the same values.}
	\label{lst:skewing-simple}
\end{figure}

\begin{figure}[!ht]
\centering
\begin{tikzpicture}
\draw[thick,->] (-1,-.5) -- (13,-.5) node[right]{$t$};
\draw[thick,->] (-1,-.5) -- (-1,5.8) node[above]{$x$};

\foreach \t in {0,1,2,3,4,5,6}
\foreach \x in {0,1,2,3,4,5}
\fill[darkgray] (\t*2,\x/2+\t/2) circle (0.1);

\foreach \t in {0,1,2,3,4,5}
\foreach \x in {1,2,3,4,5}
\draw[thick,->,gray!60] (\t*2+.2,\x/2+\t/2) -- (\t*2+1.8,\x/2+\t/2);

\foreach \t in {0,1,2,3,4,5}
\foreach \x in {0,1,2,3,4}
\draw[thick,->,gray!60] (\t*2+.2,\x/2+\t/2+.05) -- (\t*2+1.8,\x/2+\t/2+.9);

\foreach \t in {0,1,2,3,4,5,6}
\foreach \x in {0,3}
\fill[green,opacity=0.3] (\t*2-.6,\x/2+\t/2-.15) rectangle (\t*2+.6,\x/2+\t/2+1.15);

\foreach \t in {0,1,2,3,4,5,6}
\draw (\t*2,1pt-.5cm) -- (\t*2,-1pt-.5cm) node[anchor=north] {$\t$};
\foreach \x in {0,1,2,3,4,5,6,7,8,9,10,11}
\draw (1pt-1cm,\x/2) -- (-1pt-1cm,\x/2) node[anchor=east] {$\x$};
\end{tikzpicture}
\caption{A skewed iteration space. Dependencies between blocks have not changed, and interchange is \emph{not} valid. (Skewed) tiles included for reference.}
\label{fig:skewed-loops-skewed-space}
\end{figure}
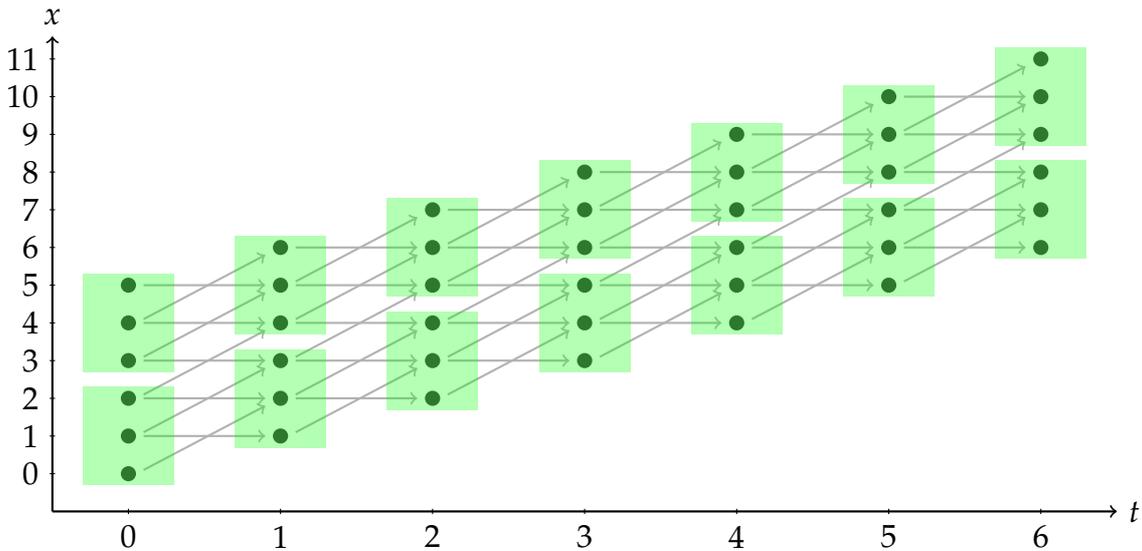

Note that skewing \emph{does not} change the execution order of the loops, nor does it change the structure of the loop nest.

\subsection{Common sub-expression elimination}
\label{sec:impl-cse}

As skewing is merely a substitution of indices and loop bounds, we do not change the expressions.
Since arithmetic on floats is generally non-associative, we especially wish to avoid changing the structure of expressions, to simplify testing of the code.

In particular, we ensure than skewing is performed \emph{before} CSE occurs, to avoid expansion during skewing.
CSE does not remove the skew, as skewing changes the indices used with reference to the loop variable (see Figure~\ref{lst:skewing-simple}), while CSE manipulates expressions.

\subsection{Implementation}
This was a straightforward substitution in the Devito symbolic engine, exactly as described above.
We iterate over the dimensions, first identifying the time dimension, then adding the offset to each incremental spatial loop variable.
We then replace every reference to the space dimension variable with a similar reference, subtracting the offset.

If there is no time dimension, no skewing is applied, as time-tiling would not be relevant.

\section{Time-tiling}
\label{sec:time-tiling}
As we noted in the previous section, skewing has not changed the execution order: when applying the tiling transformation, we now have skewed tiles on a skewed iteration space (Figure~\ref{fig:skewed-loops-skewed-space}).
We now need to `straighten' the tiles by aligning the loop bounds.
All the transformations in this section pertain to tiling, and they are performed in the Devito loop engine.

\subsection{Aligning the loop bounds}
We need to modify the tiling transformation slightly to make the interchange valid.
This was explored in Section~\ref{sec:bg-time-tiling}.
As this alignment was achieved alongside bounding for valid array accesses (detailed in the next section), Figures~\ref{lst:skew-straight} and~\ref{fig:skew-straight} are for illustrative purposes only.

\begin{figure}[!ht]
\begin{lstlisting}
for (int t_blk = t_s; t_blk < t_e; t_blk += t_blk_size)
  for (int t = t_blk; t < min(t_e, t_blk + t_blk_size); t++)
    for (int x_blk = x_s; x_blk < x_e + t_e; x_blk += x_blk_size)
      for (int x = x_blk; x < min(x_e + t_e, x_blk + x_blk_size); x++)
        A[t][(x-t)] = A[t-1][(x-t)-1] + A[t-1][(x-t)+1];
\end{lstlisting}
	\caption{Code which has been tiled following a skewing transformation. This code contains invalid array accesses.}
	\label{lst:skew-straight}
\end{figure}

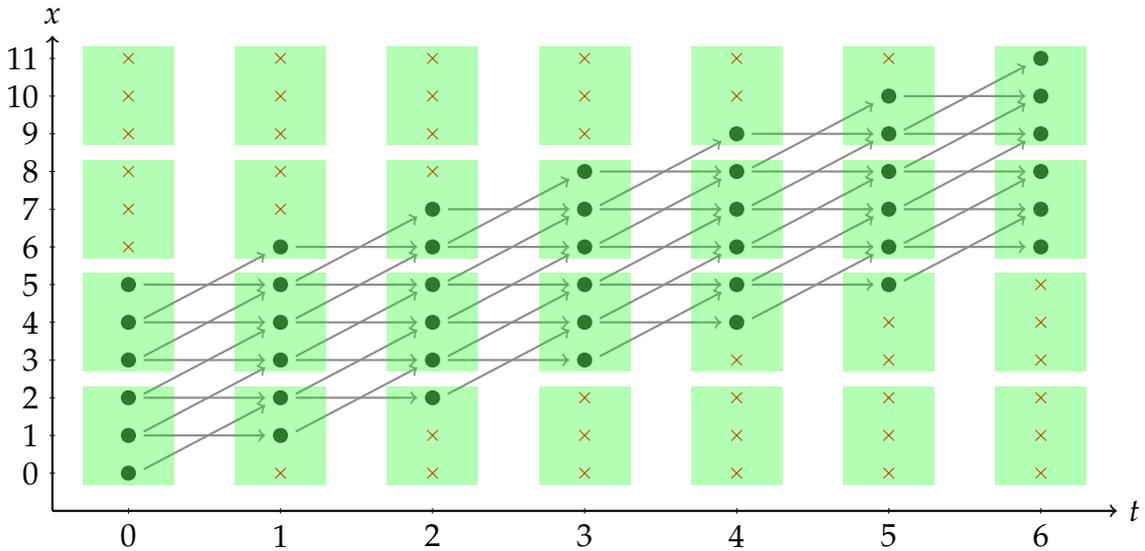
\begin{figure}[!ht]
	\centering
	\begin{tikzpicture}
	\draw[thick,->] (-1,-.5) -- (13,-.5) node[right]{$t$};
	\draw[thick,->] (-1,-.5) -- (-1,5.8) node[above]{$x$};
	
	\foreach \t in {0,1,2,3,4,5,6}
	\foreach \x in {0,1,2,3,4,5,6,7,8,9,10,11}
	\draw (\t*2,\x/2) node[cross=2.4,red]{};
	
	\foreach \t in {0,1,2,3,4,5,6}
	\foreach \x in {0,1,2,3,4,5}
	\fill[darkgray] (\t*2,\x/2+\t/2) circle (0.1);
	
	\foreach \t in {0,1,2,3,4,5}
	\foreach \x in {1,2,3,4,5}
	\draw[thick,->,darkgray!60] (\t*2+.2,\x/2+\t/2) -- (\t*2+1.8,\x/2+\t/2);
	
	\foreach \t in {0,1,2,3,4,5}
	\foreach \x in {0,1,2,3,4}
	\draw[thick,->,darkgray!60] (\t*2+.2,\x/2+\t/2+.05) -- (\t*2+1.8,\x/2+\t/2+.9);
	
	\foreach \t in {0,1,2,3,4,5,6}
	\foreach \x in {0,3,6,9}
	\fill[green,opacity=0.3] (\t*2-.6,\x/2-.15) rectangle (\t*2+.6,\x/2+1.15);
	
	\foreach \t in {0,1,2,3,4,5,6}
	\draw (\t*2,1pt-.5cm) -- (\t*2,-1pt-.5cm) node[anchor=north] {$\t$};
	\foreach \x in {0,1,2,3,4,5,6,7,8,9,10,11}
	\draw (1pt-1cm,\x/2) -- (-1pt-1cm,\x/2) node[anchor=east] {$\x$};
	\end{tikzpicture}
	\caption{Straightened tiles on a skewed iteration space, making the interchange valid. Invalid array accesses indicated by crosses.}
	\label{fig:skew-straight}
\end{figure}

\subsection{Min/max bounds for valid array accesses}
Since a skewed iteration space is not rectangular (or in general, not a hypercube), valid array accesses over a space variable will not be valid for the same values between time iterations (as seen the previous section).
As it is not feasible to generate a remainder loop for each time iteration and choice of space dimension,\footnote{This would require at least \(\sum_{m=1}^{n} \binom{n}{m} = 2^n \) remainder loop nests, where there are \(n\) spatial dimensions -- one for each choice of dimensions.} we bound the incremental loops using the \texttt{min} and \texttt{max} functions.

\begin{figure}[!ht]
\begin{lstlisting}
for (int t_blk = t_s; t_blk < t_e; t_blk += t_blk_size)
  for (int t = max(t_s, t_blk);
           t < min(t_e, t_blk + t_blk_size); t++)
    for (int x_blk = x_s; x_blk < x_e + t_e; x_blk += x_blk_size)
      for (int x = max(x_s + t, x_blk);
               x < min(x_e + t, x_blk + x_blk_size); x++)
        A[t][(x-t)] = A[t-1][(x-t)-1] + A[t-1][(x-t)+1];
\end{lstlisting}
	\caption{We add lower bounds for each of the incremental loop variables, and we change the upper bounds for the spatial dimensions to be bounded by \texttt{x\_e + t}, instead of \texttt{x\_e + t\_e}. This prevents out of bounds accesses in the lower and upper triangular regions of Figure~\ref{fig:skewing-bounded} respectively.}
	\label{lst:skewing-bounded}
\end{figure}

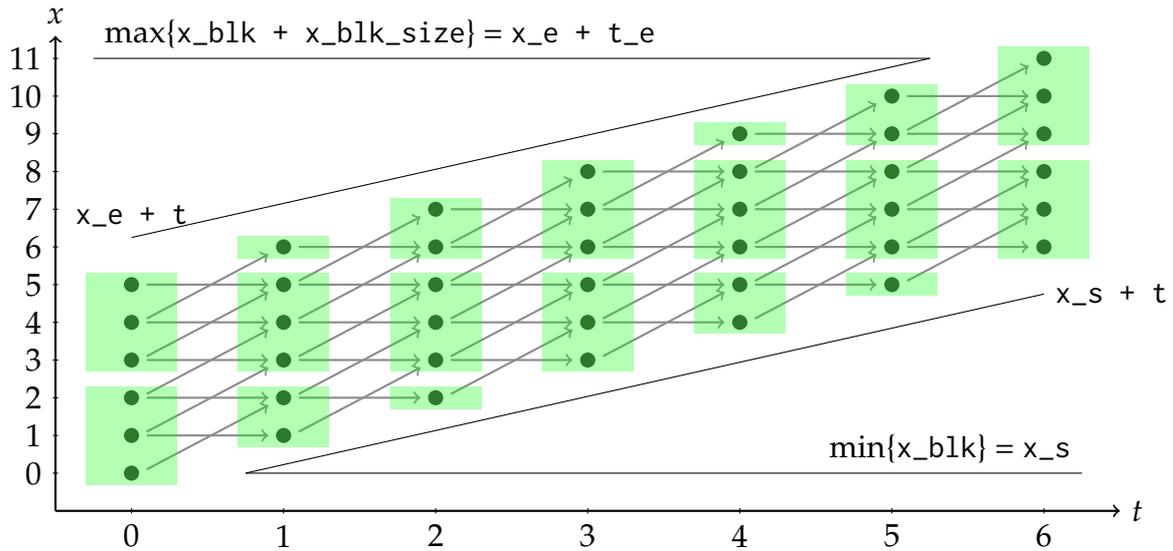
\begin{figure}[!ht]
\centering
\begin{tikzpicture}
\draw[thick,->] (-1,-.5) -- (13,-.5) node[right]{$t$};
\draw[thick,->] (-1,-.5) -- (-1,5.8) node[above]{$x$};

\draw (1.5,0) -- (12.5,0) node[anchor=south east] {min\{\texttt{x\_blk}\} = \texttt{x\_s}};
\draw (1.5,0) -- (12,2.375) node[anchor=west] {\texttt{x\_s + t}};
\draw (10.5,11/2) -- (-.5,11/2) node[anchor=south west] {max\{\texttt{x\_blk + x\_blk\_size}\} = \texttt{x\_e + t\_e}};
\draw (10.5,11/2) -- (0,3.125) node[anchor=south] {\texttt{x\_e + t}};

\foreach \t in {0,1,2,3,4,5,6}
\foreach \x in {0,1,2,3,4,5}
\fill[darkgray] (\t*2,\x/2+\t/2) circle (0.1);

\foreach \t in {0,1,2,3,4,5}
\foreach \x in {1,2,3,4,5}
\draw[thick,->,darkgray!60] (\t*2+.2,\x/2+\t/2) -- (\t*2+1.8,\x/2+\t/2);

\foreach \t in {0,1,2,3,4,5}
\foreach \x in {0,1,2,3,4}
\draw[thick,->,darkgray!60] (\t*2+.2,\x/2+\t/2+.05) -- (\t*2+1.8,\x/2+\t/2+.9);

\foreach \t in {0,1,2,3} \foreach \x in {3}
\fill[green,opacity=0.3] (\t*2-.6,\x/2-.15) rectangle (\t*2+.6,\x/2+1.15);

\foreach \t in {3,4,5,6} \foreach \x in {6}
\fill[green,opacity=0.3] (\t*2-.6,\x/2-.15) rectangle (\t*2+.6,\x/2+1.15);

\foreach \t in {0} \foreach \x in {0}
\fill[green,opacity=0.3] (\t*2-.6,\x/2-.15) rectangle (\t*2+.6,\x/2+1.15);

\foreach \t in {6} \foreach \x in {9}
\fill[green,opacity=0.3] (\t*2-.6,\x/2-.15) rectangle (\t*2+.6,\x/2+1.15);

\foreach \t in {1} \foreach \x in {1}
\fill[green,opacity=0.3] (\t*2-.6,\x/2-.15) rectangle (\t*2+.6,\x/2+.65);

\foreach \t in {2} \foreach \x in {6}
\fill[green,opacity=0.3] (\t*2-.6,\x/2-.15) rectangle (\t*2+.6,\x/2+.65);

\foreach \t in {4} \foreach \x in {4}
\fill[green,opacity=0.3] (\t*2-.6,\x/2-.15) rectangle (\t*2+.6,\x/2+.65);

\foreach \t in {5} \foreach \x in {9}
\fill[green,opacity=0.3] (\t*2-.6,\x/2-.15) rectangle (\t*2+.6,\x/2+.65);

\foreach \t in {2} \foreach \x in {2}
\fill[green,opacity=0.3] (\t*2-.6,\x/2-.15) rectangle (\t*2+.6,\x/2+.15);

\foreach \t in {1} \foreach \x in {6}
\fill[green,opacity=0.3] (\t*2-.6,\x/2-.15) rectangle (\t*2+.6,\x/2+.15);

\foreach \t in {5} \foreach \x in {5}
\fill[green,opacity=0.3] (\t*2-.6,\x/2-.15) rectangle (\t*2+.6,\x/2+.15);

\foreach \t in {4} \foreach \x in {9}
\fill[green,opacity=0.3] (\t*2-.6,\x/2-.15) rectangle (\t*2+.6,\x/2+.15);

\foreach \t in {0,1,2,3,4,5,6}
\draw (\t*2,1pt-.5cm) -- (\t*2,-1pt-.5cm) node[anchor=north] {$\t$};
\foreach \x in {0,1,2,3,4,5,6,7,8,9,10,11}
\draw (1pt-1cm,\x/2) -- (-1pt-1cm,\x/2) node[anchor=east] {$\x$};
\end{tikzpicture}
\caption{Straightened tiles on a skewed iteration space, with restrictions to valid accesses. The lower triangular reason corresponds to the \texttt{max} bound in Figure~\ref{lst:skewing-bounded}, restricting accesses by the greater of the two lines. Likewise the upper triangular reason corresponds to the \texttt{min} bound.}
\label{fig:skewing-bounded}
\end{figure}

At the same time, it makes sense for this approach to supersede the remainder loops detailed in Section~\ref{sec:remainder-loops} which generate remainder iterations.
This is valid because every array access is valid under the new bounds.
We can therefore extend the iteration space of the tile loop to the full extent of the skewed iteration space.
This would have previously resulted in out-of-bounds accesses as bounding did not occur.

Figures~\ref{lst:skewing-bounded} and~\ref{fig:skewing-bounded} combine skewing and min/max bounding.

\subsection{Parallelisation of loops}
A major requirement to maximising reductions in runtime is the parallelisation of computation.
In spatial tiling, the outer space tile loops (Figure~\ref{lst:parallel-space}) may be computed in parallel.

\begin{figure}[!ht]
\begin{lstlisting}
for (int t = t_s, t < t_e; t++)
  // parallelise the x_blk loop below
  for (int x_blk = x_s; x_blk < x_e + t_e; x_blk += x_blk_size)
    for (int x = max(x_s + t, x_blk);
             x < min(x_e + t, x_blk + x_blk_size); x++)
      A[t][(x-t)] = A[t-1][(x-t)-1] + A[t-1][(x-t)+1];
\end{lstlisting}
	\caption{Under spatial tiling, the loops iterating over the tiles may be parallelised.}
	\label{lst:parallel-space}
\end{figure}

One may hope that we can parallelise the same space tile loops under time-tiling, but this is not the case.
Consider a time tile size at least as large as the extent of the time dimension, so that the time dimension forms a single tile.
The time tile loop will only execute once, so we can ignore it (Figure~\ref{lst:parallel-time}).

\begin{figure}[!ht]
\begin{lstlisting}
// cannnot parallelise the x_blk loop, as dependences cross space tiles
for (int x_blk = x_s; x_blk < x_e + t_e; x_blk += x_blk_size)
  // parallelising the time loop is never valid
  for (int t = t_s; t < t_e; t++)
    // finally, parallelise the incremental space loops
    for (int x = max(x_s + t, x_blk);
             x < min(x_e + t, x_blk + x_blk_size); x++)
      A[t][(x-t)] = A[t-1][(x-t)-1] + A[t-1][(x-t)+1];
\end{lstlisting}
	\caption{Time-tiled loop nest (including interchange), with the outermost loop removed and bounds on the incremental time loop adjusted accordingly. The resultant loop structure is identical to spatial tiling, except with the time loop interchanged to appear within space tile loops.}
	\label{lst:parallel-time}
\end{figure}

It is clear to see that the loops iterating over the tiles cannot be parallelised, as there are dependences that cross tile boundaries.
Since parallelisation of the time loop is always invalid, the first loop that can be parallelised is the outermost space loop that iterates within a spatial tile.
This can be parallelised as it has no dependences within a time iteration.

\subsection{Implementation}
Unlike skewing implemented in the DSE, tiling was not done through a simple substitution of variables.
Instead, the visitor pattern was used to manipulate Devito's internal representation of the loop structure.

Together, the following paragraphs explain this design decision.

\paragraph{OpenMP and \texttt{fmax}}
While we have used the \texttt{max} function in examples, the only similar function in C, which Devito generates, is \texttt{fmax}.
However, it is known that we are unable to use \texttt{fmax} in a loop header when the loop is parallelised under OpenMP due to scoping and optimisation problems~\cite{openmp-fmax}.
In particular, the call to \texttt{fmax} although invariant, returns a \texttt{float} rather than an integer, and is therefore not a valid \emph{controlling predicate}~\cite{openmp-spec}.

\paragraph{Loop nest structure}
Each tiled loop would be tiled into two loops: an incremental \texttt{x} loop, and a tile \texttt{x\_blk} loop, for the tiles.
Substitution would be tricky, as the body of each loop would have to be copied and substituted in turn.
Only then could the loops be interchanged.

\paragraph{Perfect loop nests}
Definition. A \emph{perfect loop nest} is a loop which fulfilling this condition: its body contains either a perfect loop nest, or only non-loop statements.
Ordinarily, only perfect loop nests may be tiled~\cite{ahmed-imperfect}.

\begin{figure}[!ht]
\begin{lstlisting}
for (int x_blk = x_s; x_blk < x_e + t_e; x_blk += x_blk_size)
  int x_lb = fmax(x_s + t, x_blk);
  int x_ub = fmin(x_e + t, x_blk + x_blk_size);
  for (int x = x_lb; x < x_ub; x++)
    A[t][(x-t)] = A[t-1][(x-t)-1] + A[t-1][(x-t)+1];
\end{lstlisting}
	\caption{Bounds hoisted out of the loop header.}
	\label{lst:hoist-bounds}
\end{figure}

\paragraph{Direct construction of loop nests}
In Devito's existing implementation of spatial tiling, loops were tiled with a function designed to manipulate perfect loop nests.
Each loop would be duplicated, adjusting the bounds on the resultant loop, then composed, along with the body of the innermost loop.
However, as it proved impossible to insert the \texttt{fmax} function directly into the loop header, it was necessary to hoist computation of the bounds to before each loop (Figure~\ref{lst:hoist-bounds}).
As this did not resemble a perfect loop nest,\footnote{Note that the control-flow graph of the program would not change under hoisting this expression, and the code motion is legal as the values are loop-invariant.} the approach of composition failed.

\subsubsection{Visitor pattern}
Given the above constraints, we found that the visitor pattern would be the most practical solution to constructing a new tiled loop nest.
This allowed for easy propagation of loop properties, offsets to the upper and lower bounds, and a clear distinction between the parallelisable incremental loops, and the sequential tile loops.

\section{Test suite}
\label{sec:impl-tests}
Devito has an extensive test suite, which we endeavoured not to break while implementing time-tiling.
In particular, the existing tests used for spatial tiling were helpful, as they provided a benchmark for the correct tiling behaviour.

We extended the test suite, verifying the following:

\begin{itemize}
	\item Skewing must not change the numerical results.
	\item Spatial tiling behaves as it did before.
	\item Time-tiling with skewing produces the same results as non-tiled code. This would be equal to the spatially-tiled code.
\end{itemize}

\section{Auto-tuning}
\label{sec:autotune}

Time-tiling introduces two parameters that can be tuned: the skewing factor, and time tile size.

Since time-tiling had not been implemented in Devito before, it was decided that the auto-tuner should try all reasonable time tile sizes.
It was modified for this behaviour, as well as minor changes to try additional specific combinations, such as trying the maximum block size for the innermost dimension.

We have not extended the auto-tuner to try different skewing factors as we hope to draw conclusions about the skewing factors that would produce the best runtime, thus removing the need for auto-tuning in future.
Therefore, auto-tuning will be performed for the relevant manually-specified skewing factors.
Manually varying the skewing factor is not an undue burden, as there are very few plausible skewing factors.

\section{Summary and further implementation work}
\label{sec:impl-further}

In this chapter, we have provided an implementation of time-tiling for Devito, which handles perfect loop nests, the primary and best-studied scenario in time-tiling.

Below, we discuss further implementation work in time-tiling that Devito will benefit from, summarising their motivation and why we have chosen not to implement them in this project.
In Section~\ref{sec:future-work}, we will elaborate on their significance, challenges, and present some approaches to solving them.

\subsection{Source and receiver loops}
\label{sec:impl-imperfect}

Sources and receivers are commonly used with the acoustic wave equation, so that it can model not only wave propagation, but also the addition of ongoing waves or generated waves; consider sonar or tremors, to take an example from geology.
These are an important use case for Devito, as it targets the domain of seismic imaging.

However, these are added as source and receiver loops performed during each time iteration, making the time loop an imperfect loop nest.
Our implementation of time-tiling is unable to handle imperfect loop nests, although this is possible with a few more transformations~\cite{ahmed-imperfect}.
We considered this to be beyond the scope of the project, as the transformations, while only marginally more difficult than skewing conceptually, presented an equally large software engineering burden.
Further, this might present a version of Devito that diverged too far from the original, causing the basis for evaluation to be too tenuous.

We will address source and receiver loops in considerable detail later, and demonstrate that the problem is not insurmountable.

\subsection{DSE aggressive mode}
DSE aggressive mode is used to performed advanced manipulation of symbolic expressions and extract loop invariants.
Loop hoisting is a technique used in general-purpose optimising compilers to eliminate redundant computation, and this is a specific situation in which hoisting is used.

While important, we felt that supporting this feature would potentially be time-consuming and would not add significant value to our evaluation.
A specific concern is that this manipulation can change the order of floating-point operations.
Combining this with our transformations would mean that the results may no longer be numerically identical, which would hinder our evaluation, as establishing acceptable bounds on error is highly problem-dependent.
Further, it would introduce another variable to be evaluated; however, the variable of advanced expression manipulation is not immediately relevant to our time-tiling transformation.
In Section~\ref{sec:future-aggressive} we argue its consistency with our transformation, and outline a strategy for verifying and integrating the two transformations.

\subsection{Minor extensions}
\label{sec:impl-minor}

The following are fairly straightforward to comprehend, and appear accordingly straightforward to implement. We elected not to implement them in order to devote more time to a rigorous evaluation of the transformation.

\begin{description}
	\item[Skewing factor legality detection] In Section~\ref{sec:bg-skewing}, we stated the minimum legal skewing factor for time-tiling.
	Currently, the skewing factor is a user-defined parameter.
	Regardless of its source, its legality should be checked, as Devito possesses the stencil space order and can implement a safety check.
	For our evaluation, this was not necessary as we were aware of the constraints and performed extensive numerical verification.

	\item[Time buffering] Time buffering is a space-saving optimisation discussed in our evaluation, based on the observation that some data will not be required.
	At the moment, time buffering is configured to allocate the minimum amount of memory for spatial tiling; for our evaluation we have manually specified the appropriate number of time iterations to store.
	In Section~\ref{sec:time-buffering} we provide an exact bound for the size of the buffer, which should be implemented as specified.

	\item[Tiling the innermost dimension] A specific issue to be addressed with time-tiling; in our implementation, if the innermost dimension is tiled, we obtain a slower runtime by about 10\% when chosen tile size is the full extent of the dimension.
	This does not occur with our implementation under spatial tiling.
	We suggest that the combination of the floating-point \texttt{fmin} function for bounding and vectorisation could cause this unexpected behaviour, but have not investigated it in detail.
	This does not affect evaluation as in our experimentation, the auto-tuned tile sizes always use the full extent of the innermost dimension, so we have always explicitly vectorised the innermost dimension, with the complementary benefit of saving weeks of machine-time in auto-tuner trials.

	\item[Removal of the floating point \texttt{fmin} operations] Since the bounds are guaranteed to be integers, the floating-point \texttt{fmin} and \texttt{fmax} functions should be replaced with simpler comparisons that work on integers, to reduce the tile bounding overhead.

	\item[Supplementary test cases] We have already included test cases adapted from spatial tiling and ones that specifically test skewing.
	Additional test cases should be added to test the integration of time-tiling with other transformations, beyond the existing skewing and tiling tests.
	In particular, Devito could benefit from more skewing integration tests.
	This was not an issue for our evaluation, where each application of time-tiling was numerically verified to be correct.
	Finally, there are two obsolete test cases which exist to test the structure of remainder loops, which have been superseded by our loop bounding strategy.
	These should be deprecated and eventually removed.

	\item[Removal of zero-iteration tiles] These are tiles which contain no iterations as they are outside the boundary of the skewed iteration space; those occurring in the triangular region of Figure~\ref{fig:skewing-bounded}.
	The polyhedral compiler CLooG removes these at the cost of an additional bounding computation per dimension per time tile.
	Clearly a trade-off is to be made here: it may only be worthwhile performing this when the skewing factor is large.
\end{description}

	\documentclass[thesis.tex]{subfile}

\chapter{Evaluation}
\label{ch:evaluation}

This chapter is devoted to the performance evaluation of our implementation of time-tiling in Devito.
It is organised into a number of sections:

\begin{description}
	\item[Section~\ref{sec:eval-objs}] We recall the motivation of this project, and the objectives of our evaluation.

	\item[Section~\ref{sec:time-tiling-details}] The restrictions that time-tiling imposes on our use of Devito, and the considerations that it warrants.

	\item[Section~\ref{sec:arithmetic-intensity}] A novel estimator for \emph{arithmetic intensity under time-tiling}, including a proof of its validity and an algorithm to understand ideal cache usage under time-tiling.

	\item[Sections~\ref{sec:test-method}, \ref{sec:roofline-intro}, and~\ref{sec:eval-params}] Our testing methodology, and an introduction to the \emph{roofline model}, which is widely used to understand the performance of stencil calculations.
	Additionally, the parameters varied in our tests, and the reasoning behind them.

	\item[Sections~\ref{sec:perf-laplace} and~\ref{sec:perf-awe}] Performance results from two families of stencils, generated by the Laplace and acoustic wave equation operators respectively.

	\item[Sections~\ref{sec:eval-skewing-effect} and~\ref{sec:tt-size-effect}] An analysis of the effect of the skewing factor and time tile size respectively on runtime of a stencil computation under time-tiling.

	\item[Section~\ref{sec:ai-eval}] An evaluation of the accuracy of our arithmetic intensity estimator, the circumstances under which it is most useful, a model for when performance improvement declines, and a comparison to the existing estimator in Devito, with discussion of areas for improvement.

	\item[Section~\ref{sec:further-eval}] Limitations of our performance testing and further work in evaluation.
	
	\item[Section~\ref{sec:eval-conclusion}] A summary of this chapter.
\end{description}

\section{Objective}
\label{sec:eval-objs}

The motivation for this project is to reduce computational runtime by exploiting data locality between time iterations through time-tiling.
We have described the motivation and investigated tools which have achieve performance gains from the technique.
It had previously been shown that Devito benefits from reductions in runtime of up to 27.5\% with time-tiling using CLooG~\cite{dylan}.

Chapter~\ref{ch:implementation} described our implementation of time-tiling in Devito.
The objectives of this evaluation are as follows:

\begin{itemize}
	\item Measure the performance gain against non-tiled and spatially-tiled computations;
	\item Verify any requirements to perform time-tiling effectively;
	\item Determine procedures and heuristics to identify the best-performing parameters for time-tiling;
	\item Additional verification of correctness, and margin of error introduced against non-tiled code\footnote{For example, due to non-associativity of floating-point arithmetic.} if any.
\end{itemize}

\section{Details for evaluating time-tiling}
\label{sec:time-tiling-details}

As time-tiling has been implemented within Devito, there is no need for any external tool such as a polyhedral compiler.
Nevertheless, the following topics warrant careful consideration when applying and analysing time-tiling.

\subsection{Time-buffering}
\label{sec:time-buffering}

Time-buffering is a memory-saving technique.

\begin{figure}[!ht]
\begin{lstlisting}
for (int t = t_s; t < t_e; t++)
  for (int x = x_s; x < x_e; x++)
    A[t][x] = A[t-1][x] + A[t-2][x] + ... + A[t-n][x];
\end{lstlisting}
	\caption{A stencil with a value depending on data in the previous \(n\) time iterations. \(n\) is usually small compared to the problem domain, (\texttt{t\_e - t\_s + n}) here.}
	\label{lst:buffer-eg}
\end{figure}

A stencil may only have dependencies on data from finitely many previous time iterations.
Consider a stencil with a data dependence on the last \(n\) time iterations (Figure~\ref{lst:buffer-eg}).
It is clear that whenever \(t > n\), data from the 0-th time iteration is no longer required, and can be overwritten.

When space-tiling in Devito, each time iteration only begins once the previous iteration has finished.
Therefore, we only need storage for \(n+1\) time iterations of data: the current iteration being computed; and the previous \(n\) iterations, its dependencies.

As we have not implemented the calculation for a correct time buffer size in Devito, we did not use time buffering for the Laplace operator stencils (Section~\ref{sec:perf-laplace}), and manually specified a buffer size for the acoustic wave equation operator stencils (Section~\ref{sec:perf-awe}).
Nevertheless, we can establish bounds on the size of a time buffer: \(n+t\) time iterations must be stored, where \(t\) is the size of a time tile.\footnote{Following the logic used for spatial tiling, we need \(n\) time iterations of dependencies and \(t\) iterations of the time tile being computed (1 in the case of spatial tiling).}

\subsection{Auto-tuner iterations}
Previously covered in Section~\ref{sec:autotune}, the Devito auto-tuner experimentally searches for tile sizes achieving the lowest runtime.
To do this, it needs to store some time iterations of computations: this value is governed by a parameter.

For its results to be meaningful, we need to compute enough time iterations to distinguish between large time tile sizes.
Without time-tiling, a fairly small number of time iterations (4) sufficed for auto-tuning.
In our evaluation, we decided that the largest desirable time tile size was 16, to which we set the parameter.

\subsection{Time-tiling parameters}
Time-tiling provides more parameters: a valid skewing factor, and a tile size for the time dimension.
The tile size search was integrated into the auto-tuner, as it already had functionality to search for spatial tile sizes.

It was decided to try skewing factors by hand, as their validity would depend on the stencil (Section~\ref{sec:bg-skewing}), and we wished to find a heuristic for the skewing factor resulting in the lowest runtime.
We have found that the minimum skewing factor produces the lowest runtime (Section~\ref{sec:eval-skewing-effect}).

\section{Arithmetic intensity under time-tiling}
\label{sec:arithmetic-intensity}

Recall that \emph{arithmetic intensity} is defined to be the number of floating point operations performed per byte.
Time-tiling is an optimisation to improve the reuse of cached data, effectively increasing the arithmetic intensity of a stencil, as data needs to be fetched from memory less frequently.

\begin{framed}
In this section, \(d_t\) represents the time dimension and \(d_i (i=1,2,\cdots)\) represent a spatial dimension;
\(D_i\) represents the extent of dimension \(d_i\), or \(D_t\) the extent of the time dimension;
\(\delta_i\) the maximum spatial dependence (radius) of the stencil in dimension \(d_i\), \(\delta_t\) the distance of the time dependence;
\(t_i\) the chosen tile size in dimension \(d_i\), \(t_t\) for the time tile size;
and \(\Omega\) represents the size of the cache.
\end{framed}

\subsection{Arithmetic intensity estimator in Devito}
Suppose a stencil requires data from the last \(\delta_t\) time iterations.
Devito reports arithmetic intensity as `operational intensity', and calculates it as follows:

\begin{description}
	\item[Non-tiled code] Arithmetic intensity is calculated from the definition, assuming that data that has been loaded into the cache would \emph{not} need to be re-loaded in the same time iteration.
	In reality, with a sufficiently large iteration space, each byte would have to be loaded into cache more than once for each time iteration.
	The measure therefore \emph{overestimates} the number of floating point operations per byte loaded from memory and hence true arithmetic intensity.
	In the graphs later in this chapter, an overestimate results in a \emph{right shift} of a data point from its true location.

	\item[Space-tiled code] Again, Devito assumes data need only be loaded into the cache once per time iteration, hence would be used fully before being evicted, calling this scheme `compulsory memory traffic'.
	For each time iteration, this requires the previous \(\delta_t\) time iterations of data to be loaded from memory.
	This produces the same figure for arithmetic intensity as the measure for non-tiled code above.
	Since the dependencies of tiles overlap in previous time iterations, with a sufficiently large iteration space, some dependencies will need to be loaded into cache more than once (assuming at least two spatial dimensions with non-zero space order).\footnote{This is fairly simple to see. Ignore all but two spatial dimensions with non-zero space order, and suppose otherwise for a contradiction. Arbitrarily choose a finite cache size of \(\Omega\) floating-point numbers, and call our spatial dimensions \texttt{x}, \texttt{y} with dimension extents \(x, y\) respectively. W.l.o.g.~\(x \le y\) and that the stencil has space order \(\delta>0\) in both dimensions (take the minimum). Now we need \(x\delta \le \Omega\), since minimising dependency overlap along the \texttt{x} dimension demands that we calculate the first \texttt{x} `row' before any other \texttt{x}-tiles. But we can choose \(x > \Omega / \delta\).}
	Thus this measure also \emph{overestimates} arithmetic intensity, but to a smaller extent than with non-tiled code.
	Later, we explore the possibility of a tighter bound.
\end{description}

Remark: later, we will see that it is preferable to overestimate the arithmetic intensity of a stencil computation than underestimate it, and we provide arguments for why the measures stated in this section are useful and fairly realistic.
This will be explained in detail when studying the roofline model in Section~\ref{sec:roofline-intro}, as it requires other ideas that have yet to be introduced.

\subsection{Estimating arithmetic intensity under time-tiling}
We need to establish a new measure for arithmetic intensity under time-tiling, preferably one compatible with the existing measure under spatial tiling.
We consider the following statements to follow naturally from Devito's calculation of arithmetic intensity under spatial tiling.

\begin{itemize}
	\item Analogously to space-tiling, we assume that data will be used fully before being evicted from cache.
	\item If \(t_t\) is the time tile size, we can calculate \(t_t\) time iterations of data from \(n\) previous time iterations loaded into the cache, as opposed to calculating 1 time iteration from the same data.
	\item Thus we have performed \(t_t\) times as many floating-point operations on it.
\end{itemize}

A natural estimator for the arithmetic intensity of a stencil computation under time-tiling might be the arithmetic intensity of the same computation under spatial tiling multiplied by the size of the time tile used.

This is consistent with the measure under spatial tiling, since a spatially tiled loop is similar a time-tiled version of the loop under a (legal) interchange of the time and spatial tile loops.
This would also result in an overestimate; the next section explores the magnitude of this overestimate under spatial and time-tiling.

We will refer to this estimator as the \emph{naive estimator}, which only requires information about the stencil.
Now we will explore tighter bounds assuming ideal scheduling; later in our analysis (Section~\ref{sec:ai-eval}) we will demonstrate why they are necessary.

\subsection{Tile boundaries determine cache reuse}
Recall that in spatial tiling, we contrive small iteration spaces (`tiles') such that the data dependencies from the previous time iteration could fit within the cache.
Therefore, the internal volume of a tile need only be transferred from memory once (Figure~\ref{fig:tile-reload}).
However, there is no such guarantee for the boundary region of a tile.

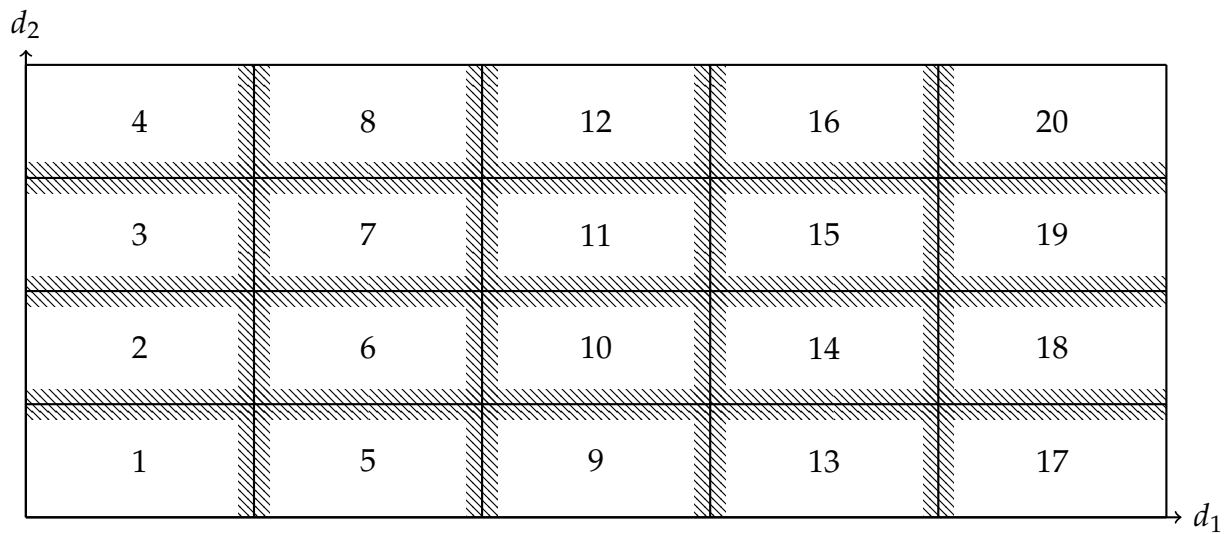
\begin{figure}[ht]
	\centering
	\begin{tikzpicture}
	\draw[xstep=3cm,ystep=1.5cm,black,thick] (0,0) grid (15,6);

	\foreach \x in {1,2,3,4}
		\fill[pattern=north west lines,pattern color=black] (\x*3-.2,0) rectangle (\x*3+.2,6);
	\foreach \y in {1,2,3}
		\fill[pattern=north west lines,pattern color=black] (0,\y*1.5-.2) rectangle (15,\y*1.5+.2);
	
	\foreach \x in {0,1,2,3,4}
	\foreach \y in {1,2,3,4}
		\pgfmathsetmacro\seq{\x * 4 + \y}
		\node at (3*\x+1.5,1.5*\y - 0.75) {\pgfmathprintnumber{\seq}};

	\draw[thick,->] (0,0) -- (15.2,0) node[right]{\(d_1\)};
	\draw[thick,->] (0,0) -- (0,6.2) node[above]{\(d_2\)};
	\end{tikzpicture}
	\caption{Tiles over a 2D iteration space. The clear regions need only be loaded once, but data in the hatched regions may need to be loaded multiple times, depending on the cache size. We will call the hatched region the \emph{tile boundaries}.}
	\label{fig:tile-reload}
\end{figure}

Consider a (2-dimensional) tiling in which we iterate over the \(d_1\)-tiles, then the \(d_2\)-tiles, or as the order indicated in Figure~\ref{fig:tile-reload}.
In this situation, the tile boundaries between \(d_2\)-tiles such as 1, 2 may fit within the cache; if so it is clear that none of these boundaries will need to be loaded twice (apart from the intersections with the \(d_1\)-boundaries).
Call this, the common boundary between two iterations of a \(d_2\)-tile loop, a \(d_2\)-face.
Similarly, if the boundary between \(d_1\)-rows fits within the cache, such as the boundary between tiles 1--4 and 5--8, then it is a possibility that the data within these boundaries need not be loaded twice.
Likewise, call this boundary a \(d_1\)-face, and in general call the common boundary between two iterations of the \(d_i\)-tile loop a \(d_i\)-face, with size \(F_i\).

Now consider the \(n\) dimensional case, first iterating over the \(d_1\)-tiles, then the \(d_2\)-tiles, and so on.
We make the following claim:

\begin{framed}
	If \(i \le n\) and \(F_{i} \ge \Omega\), then data of equivalent size to the contents of \emph{every} \(d_j\)-face must be loaded into the cache at least twice, for \(1 \le j < i\).
\end{framed}

For proof, we demonstrate the fact for the \(d_{i-1}\)-faces over the \(d_{i-1} d_i\)-plane; the rest by backwards induction.

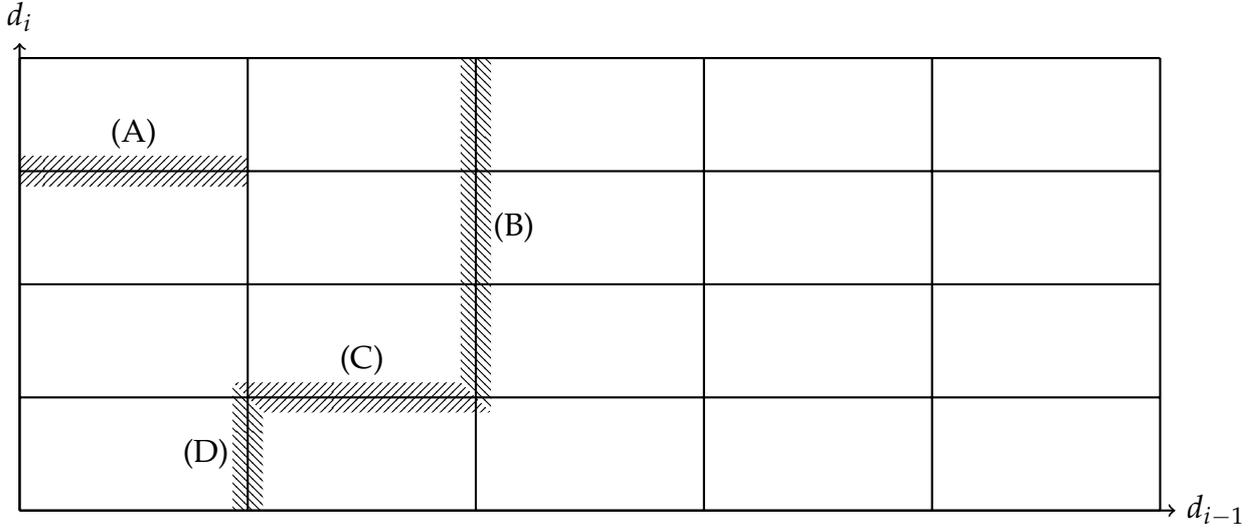
\begin{figure}[ht]
	\centering
	\begin{tikzpicture}
	\draw[xstep=3cm,ystep=1.5cm,black,thick] (0,0) grid (15,6);

	\fill[pattern=north east lines,pattern color=black] (0,4.3) rectangle (3,4.7);
	\node at (1.5,5) {(A)};

	\fill[pattern=north west lines,pattern color=black] (5.8,1.7) -- (5.8,6) -- (6.2,6) -- (6.2,1.3);
	\node at (6.5,3.75) {(B)};
	
	\fill[pattern=north east lines,pattern color=black] (2.8,1.7) -- (5.8,1.7) -- (6.2,1.3) -- (3.2,1.3);
	\node at (4.5,2) {(C)};

	\fill[pattern=north west lines,pattern color=black] (2.8,0) -- (2.8,1.7) -- (3.2,1.3) -- (3.2,0);
	\node at (2.45,0.75) {(D)};

	\draw[thick,->] (0,0) -- (15.2,0) node[right]{\(d_{i-1}\)};
	\draw[thick,->] (0,0) -- (0,6.2) node[above]{\(d_i\)};
	\end{tikzpicture}
	\caption{Tiles in the \(d_{i-1} d_i\)-plane. We consider iterations crossing the \(d_2\)-faces corresponding to regions A and C.}
	\label{fig:tile-zigzag}
\end{figure}

Suppose that a \(d_i\)-face (region A in Figure~\ref{fig:tile-zigzag}) cannot fit into the cache, i.e.~\(F_i > \Omega \).
Observe that region C (opposing hatch pattern to B, D) has the same size as region A, \(F_i\), and suppose some of the contents of a \(d_{i-1}\)-face (equal to regions B + D) do not need to be loaded into cache, and that we are evaluating the \(d_i\) tile across the C-boundary.
But region C is larger than the cache, so we cannot accommodate C alongside any points from B, D.
Thus data equivalent to the combined areas of B, D must be loaded from memory at least twice, and the claim follows.

\subsection{Establishing a tighter bound}
Let \(\bb_i\) be the union of \(d_i\)-faces.
It is a fact that the \(d_i\)-faces are distributed proportionally across the \(d_i d_j\)-plane whenever \(i \ne j\); in particular they are distributed evenly across the iteration space:

\[ \frac{\left| \bb_i \cup \bb_j \right|}{\prod_{k=1}^n D_k} = \frac{\left| \bb_i \right| + \left| \bb_j \right|}{\prod_{k=1}^n D_k} - \frac{\left| \bb_i \right| \times \left| \bb_j \right|}{\left(\prod_{k=1}^n D_k\right)^2} \]

All that remains is to find expressions for \(F_i, \left|\bb_i\right|\). Have:

\[ F_i = 2 \left(\prod_{j=1}^{i-1} t_j \right) \delta_i \left(\prod_{k=i+1}^{n} D_k \right) \]

since we need to account for the full extent of each lower dimension in the loop structure, the width of overlap in the \(i\)-th dimension (\(2 \delta_i\)), and the tile sizes in the higher dimensions; and:

\[ \frac{\left|\bb_i\right|}{\prod_{k=1}^n D_k} = 2 \frac{\delta_i}{D_i} \times \floor*{\frac{D_i}{t_i}} \]

as \(2\sfrac{r_i}{D_i}\) is the proportion of each \(d_i\)-face to \(D_i\), and \(\floor*{\sfrac{D_i}{t_i}}\) is the number of \(d_i\)-faces.

Finally, let \(\bb = \bb_1 \cup \bb_2 \cup \cdots \cup \bb_{i-1}\) for \(F_i \ge \Omega > F_{i-1} \) whenever \(1 < i < n\); if there is no such \(i\), then let \(\bb = \varnothing \).
Then divide the original arithmetic intensity estimate by the following to get a tighter bound:

\[ 1 + \frac{|\bb|}{\prod_{k=1}^n D_k} \]

Note that this method will only yield a tighter bound if \(F_i \ge \Omega\) when \(i > 1\).
If only \(F_1 > \Omega \), we can use a variation of this algorithm.
We have that for every \(d_1\)-face, \(F_1 - \Omega\) points do not fit into the cache, and have \(\floor*{\sfrac{D_1}{t_1}}\) \(d_1\)-faces.
Then consolidating these points we use:

\[\left|\bb\right| = \left(F_1 - \Omega\right) \times \floor*{\frac{D_1}{t_1}} \]

\subsection{Generalisation to time-tiling}
As a summary, the previous sections gave an algorithm to find additional memory transfer on tile boundaries within a single time iteration, under spatial tiling.
However, within a time tile, \(t_t\) time iterations are first completed.
This increases the required storage for any spatial \(d_i\) face by \(t_t\) times.
Therefore, proceed as above, except multiplying each \(F_i\) by \(t_t\).

Do note that the naive spatial tiling estimator is first multiplied by the time tile size to establish the naive time-tiling estimator.

We will use the above estimators for arithmetic intensity under both spatial and time-tiling.
While this has the disadvantage of potentially decreasing the arithmetic intensity of a stencil under spatial tiling from non-tiled code, we will not be analysing the performance of non-tiled code in relation to the roofline model detailed in Section~\ref{sec:roofline-intro}.

To summarise, the required data to calculate the arithmetic intensity under time-tiling are:

\begin{enumerate}
	\item Extents of the iteration space (\(D_t, D_1, D_2, \cdots\))
	\item Stencil spatial and time dependences (\(\delta_t, \delta_1, \delta_2, \cdots\))
	\item Chosen tile sizes (\(t_t, t_1, t_2, \cdots\))
	\item Size of the cache (\(\Omega\))
\end{enumerate}

Note that due to space constraints, we have not reported the auto-tuned tile sizes in our tables giving the arithmetic intensity data.
This data, along with complete logs of program output and auto-tuner results, is fully available in the project archive.

\subsection{Improving the bounds further}
Note that the spatial tiling bound given can be improved further, as we did not consider need to partially re-load data from the \(d_i\)-faces.
This would have a corresponding effect on the time-tiling bound.
However, we felt this would add significant complication for meagre benefit.

Additionally, note that the cache would (at times) need to not only contain a \(d_i\)-face, but also \(d_{i+1}, \cdots d_{n}\)-faces, by a similar argument to the one presented with Figure~\ref{fig:tile-zigzag}.
Therefore it is sufficient to fulfil \(F_i + F_{i+1} + \cdots + F_n > \Omega\).

However, since the extent of any dimension is likely to dwarf any tile size, this is unlikely to be relevant; in practice with 3 spatial dimensions, the auto-tuned tile sizes mean that only the \(F_1 > \Omega \) case is relevant.
The method is nevertheless crucial for the generalisation to time-tiling.

\subsection{Minimising runtime based on tile size choices}
The objective of tiling is to improve cache reuse.
Therefore, we wish to reduce the extent to which tile-face boundaries exceed the cache, where possible.

Since a stencil and problem domain will fix all values except for the \(t_t, t_i (i=1,2,\cdots)\), we are only free to choose tile sizes; in this model, tile sizes should be chosen to minimise \(|\bb|\).
This could be done through linear programming, or trivially, as an extension of the Devito auto-tuner.
Later, we assess the effectiveness of our model in predicting runtime decrease.

\section{Testing methodology}
\label{sec:test-method}

Realistic test cases vary from those that are more memory intensive to those that are more computationally intensive on a given hardware configuration.
Under the former regime, computational (arithmetic) intensity is fairly low, and the CPU uses data faster than it can be transferred from memory; in the latter case, data can be transferred more rapidly than it can be utilised.

A key premise of Devito is that arithmetic intensity can be decreased at the cost of higher memory pressure, by manipulating expressions~\cite{fabio-memory}.
Time-tiling reduces memory pressure by increasing reuse of cached data before it is evicted.\footnote{However, this effectively increases the arithmetic intensity, as discussed in this chapter.}
Accepting the premise that arithmetic intensity can be reduced sufficiently, time-tiling can be used to full effect in reducing memory pressure.\footnote{See Section~\ref{sec:roofline-intro} (``Roofline model'') for more detail.}
Therefore, the most relevant test cases to time-tiling are those which are bound by memory throughput.

\subsection{Hardware and software environment}
Evaluation was performed on a machine equipped with a single Intel Xeon\textregistered\ E5-2470 operating at 2.30 GHz, with 8 cores and 16 threads, 20MB of L3 cache, and 64 GB of DRAM.
It ran Ubuntu 16.04 LTS, with all running services required either for the operating system or our evaluation, to minimise external effects on runtime.

To ensure a realistic testing environment and utilise all the available resources, the following hold throughout the experiments:

\begin{itemize}
	\item OpenMP\footnote{OpenMP is a standard API used for compilers to implement parallelism~\cite{openmp-spec}.} directives enabled in Devito. Indicates that a loop should be computed in parallel (typically the body of the incremental time loop).
	\item OpenMP environment variable to utilise all 16 available threads.
	\item OpenMP environment variable not to migrate threads (``thread pinning''), as well as allocate threads to different cores. Thread migration incurs a significant performance overhead, and parallelism would not help if all threads executed in serial on one core.
	\item Use of the Intel C Compiler, \texttt{icc17}, enabling parallelism and the highest level of optimisation through Devito's invocation.
\end{itemize}

\subsection{Use of auto-tuner}
We wanted to experiment with tile sizes optimal under time-tiling, as they are not easily predicted from those optimal for spatial tiling or cache size~\cite{lam91}.
Therefore, the extended Devito auto-tuner (Section~\ref{sec:autotune}) was used to explore as many plausible tile sizes combinations as possible.

As detailed above, experimentation on the skewing factor was done manually.
An exhaustive search was performed, as the space was relatively small.
Further, finding a heuristic for optimal skewing factors is an objective of our evaluation.

\subsection{Functional correctness}
\label{sec:eval-func-corr}

In addition to new test cases, to build confidence in correctness of the newly-implemented skewing and tiling transformations, each application of time-tiling was numerically verified against a non-tiled computation.

In all experiments, the results were discovered to be equal.\footnote{We refer to equality under floating-point comparison; this is \emph{bitwise equality}.}
Since floating-point arithmetic is non-associative, the natural implication is that the resulting value in each field was reached through the same expressions, although the computations had been re-ordered.
Additional checks were used to determine that the results were not merely zeroes, or diverging to infinity.

\subsection{Experimentation and reporting of runtimes}
Wherever runtimes have been collected in this evaluation, the reported figure will be the \emph{minimum} of the runtimes collected in trials.
This is taken as most representative, as any noise on the testing machine is minimised, since outside factors can only increase the runtime of our computations.
In additional, the variance of the minimum is much lower than the variance of the mean in our experimentation.

Using the Laplace stencils, the auto-tuner was run for at least three trials.
The most tile size combination with the overall lowest runtime was selected, and run for a further seven trials, thus we report the minimum of 10 trials.
This obviated the need to run auto-tuning for each data point, reducing the time needed to obtain a data point by 65\%, or about a week of machine time.

For the acoustic wave equation stencils, we used the auto-tuner for each of the 10 trials run for each data point.
Thus the reported figure is the minimum of 10 auto-tuned trials.

\section{Roofline model}
\label{sec:roofline-intro}

The roofline model describes how arithmetic (or \emph{operational}) intensity kernel affects its performance on a given system.
In particular, it gives upper bounds for performance based on memory bandwidth and CPU performance, and describes the bottlenecks that a program would encounter based on its arithmetic intensity~\cite{roofline}.
The model states that as arithmetic intensity increases, performance increases at a rate determined by the memory bandwidth, until it reaches the performance limit (in Flop/s, floating-point operations per second) of the processor.

\subsection{Bounds of the test machine}
The Sandy Bridge architecture supports execution of up to 16 single-precision floating point operations per cycle, giving a theoretical maximum of 294.4 GFlop/s.\footnote{Calculated as 16 floating-point operations/cycle \(\times 2.3 \times 10^9\) cycles/sec/core \(\times\) 8 cores (on a single-node machine) \(= 294.4\) GFlop/s.}
The \emph{LINPACK} benchmark~\cite{linpack}, highly optimised and likely to perform faster than any stencil we evaluate, achieves a maximum of 131 GFlop/s on this machine; it is standard practice to double this figure to obtain the maximum for single-precision floating-point operations, giving a practical bound of 262 GFlop/s in our test scenarios.
In practice, few programs begin to approach the performance achieved by the \emph{LINPACK} benchmark.
The \emph{STREAM} benchmark~\cite{stream} indicates that the peak memory bandwidth of the machine is 17.3 GB/s.

\subsection{Arithmetic intensity and cache reuse}
Figure~\ref{fig:roofline-benchmark} shows and explains the bounds of a stencil of a given arithmetic intensity.
A given stencil computation can be plotted as a point on the graph, determined by its GFlop/s achieved, which we measure at runtime, and its inherent arithmetic intensity.
Clearly, these are strict bounds on the performance of a stencil.

\begin{figure}[!ht]
\centering
\begin{tikzpicture}
\begin{axis}[
legend style={at={(0.95,0.05)},anchor=south east},
xlabel={Arithmetic intensity},
ylabel={Performance (GFlop/s)},
domain=0.4:96,
xmode=log,
log basis x={2},
ymode=log,
log basis y={2},
log ticks with fixed point,
]
\addplot [domain=0.46:15.15,name path=A] {17.3*x};
\addplot [domain=15.15:80] {262.01};
\addplot +[mark=none,dashed,thick,name path=X] coordinates {(15.15, 8) (15.15, 262.01)};
\addplot[gray, pattern=north west lines] fill between [of=A and X];
\end{axis}
\end{tikzpicture}
\caption{Roofline graph for the test machine. Performance of any stencil (GFlop/s) of given arithmetic intensity cannot exceed the solid lines plotted. A stencil falling into the hatched region (arithmetic intensity < 15.15) will be bounded by memory bandwidth indicated by the sloped line, and those in blank region (arithmetic intensity > 15.15) are bounded by processor performance, the horizontal line. These two solid lines are the \emph{roofline}.}
\label{fig:roofline-benchmark}
\end{figure}
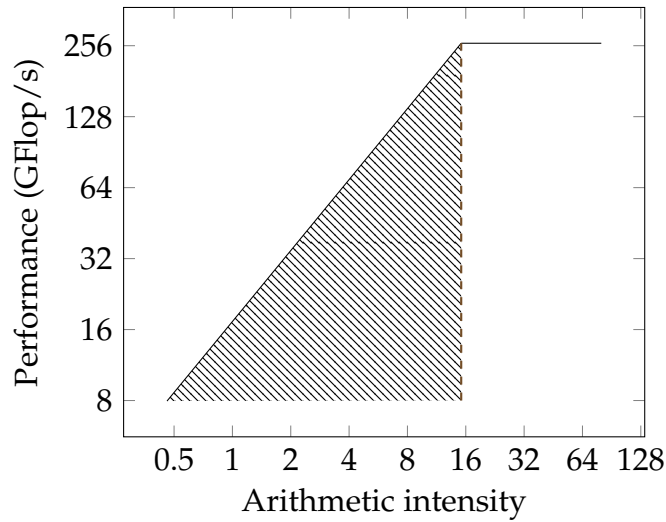

We previously suggested that it was preferable to overestimate arithmetic intensity than underestimate it.
As we can see from the figure, if we permit ourselves to underestimate arithmetic intensity, a point could lie \emph{above} the roofline, contradicting the model.
Therefore, we must not underestimate arithmetic intensity.

Recall that the objective of time-tiling is to maximise cache reuse by exploiting data locality between time iterations.
As previously argued, if a computation is bounded by memory throughput (hence in the shaded region), improving effective throughput through more effective cache usage could be reasonably argued to decrease the runtime.
From the previous discussion (Section~\ref{sec:arithmetic-intensity}), we established that tiling transformations effectively increase the arithmetic intensity of a stencil computation, resulting in a rightward shift of a data point on this graph.

We see from the diagram that such a rightward shift starting from any point in the shaded region increases the performance bound under the roofline.
Therefore, we seek to demonstrate that this is accompanied by an upward shift (together a diagonal shift) of the performance of a memory-bounded stencil computation.
In particular, we hypothesise that the performance increase of stencils that are more memory bounded (lower arithmetic intensity) will yield the greatest benefit from time-tiling.

\section{Test parameters}
\label{sec:eval-params}

\paragraph{Skewing factor}
The skewing factor was varied to understand the relationship between skewing factor and runtime.
In particular, we wished to determine if the minimum legal skewing factor would result in the minimum runtime as hypothesised.

\paragraph{Space order}
The space order of the computation determines the precision of the result.
In solving differential equations, this is the order of the approximation, beyond which smaller terms are ignored.
A higher space order results greater precision, stencil radius, and minimum skewing factor, and results in a higher arithmetic intensity, which we study.

\paragraph{Time tile size}
For each operator, we report two sets of figures: a set of auto-tuned results with any time tile size and varying skewing factors; and a set with varying time tile sizes, to inform separate analyses on skewing factors and time tile sizes.

\paragraph{Domain size}
The size of the domain is the product of all four dimensions in the simulation.
As memory was limited to 64GB needed to store both time-tiled and non-tiled results for numerical verification, we decided to use shorter time dimensions, maximising the spatial dimension size for realism.
We justify this choice reasoning that runtime should be proportional to the number of time tiles executed.\footnote{This was verified using smaller spatial dimensions.}
Therefore, we chose time dimensions 16 or 32, as these are small multiples of the time tile size.
To ensure the results are valid, we need to ensure that a single time iteration well exceeds the size of the cache.

\section{Performance of the Laplace operator}
\label{sec:perf-laplace}

\subsection{Application of the operator}
The Laplacian is the operator giving the divergence of the gradient of a scalar function, commonly used in mechanics.
In our evaluation, we used three spatial dimensions and a time dimension, with a deterministically generated input domain for numerical verification.
The input domain was chosen to avoid divergence of values, again for the purposes of numerical verification.

We previously stated that we were studying a family of stencils generated by this operator.
Devito makes it straightforward to generate stencils of varying space order, otherwise a non-trivial task; we have considered the Laplace operator with space orders 2, 4, 8, and 16.

\subsection{Results}
Two grid sizes, \(32 \times 500^3 \) and \(16 \times 600^3 \), were chosen to comply with our memory limit.
These represent 32 time iterations over a grid with 500 points in each spatial dimension and 16 time iterations over a grid with 600 points in each spatial dimension respectively.
Table~\ref{tab:laplace-results} shows the runtimes arising from running time-tiling compared to spatial tiling under the Laplace operator.
Figure~\ref{fig:laplace-graph} provides the corresponding graph for the \(16 \times 600^3 \) grid; the results and graphs for both grids are similar.

\begin{table}[p]
\centering
\begin{tabular}{rr|cccc|cccc}
\toprule
& Grid size & \multicolumn{4}{c}{\( 32 \times 500^3 \)} & \multicolumn{4}{c}{\( 16 \times 600^3 \)} \\
& Space-order & 2 & 4 & 8 & 16 & 2 & 4 & 8 & 16 \\
\midrule
N & \footnotesize runtime (s) & 6.121 & 7.572 & 10.65  & 16.941 & 5.189 & 6.402 & 9.026 & 14.87 \\
S & \footnotesize runtime (s) & 3.964 & 3.760 & 4.060 & 5.778 & 3.489 & 3.376 & 3.684 & 4.980 \\
T & \footnotesize runtime (s) & 2.939 & 2.957 & 3.762 & 5.718 & 2.710 & 2.616 & 3.288 & 4.814 \\
& \footnotesize decrease (\%) & 25.9\% & 21.4\% & 7.34\% & 1.04\% & 22.3\% & 22.5\% & 10.8\% & 3.33\% \\

\midrule
(t,8) & \footnotesize runtime (s) & 3.348 & 3.257 & 3.762 & 5.718 & 3.089 & 2.966 & 3.341 & 4.814 \\
(t,4) & \footnotesize runtime (s) & 3.114 & 3.090 & 3.764 & - & 2.844 & 2.811 & 3.288 & - \\
(t,2) & \footnotesize runtime (s) & 2.956 & 2.957 & - & - & 2.788 & 2.616 & - & - \\
(t,1) & \footnotesize runtime (s) & 2.939 & - & - & - & 2.710 & - & - & - \\
\bottomrule
\end{tabular}
\caption{Runtimes from using the Laplace operator, space orders 2, 4, 8, 16. N, S, T represent minimum runtimes without tiling, with spatial tiling, and with time-tiling respectively. (t,\(k\)) indicates results for time-tiling with a skewing factor of \(k\); T indicates the minimum of these. `-' denotes invalid skewing factor. Decrease is from spatial tiling to time-tiling.}
\label{tab:laplace-results}
\end{table}

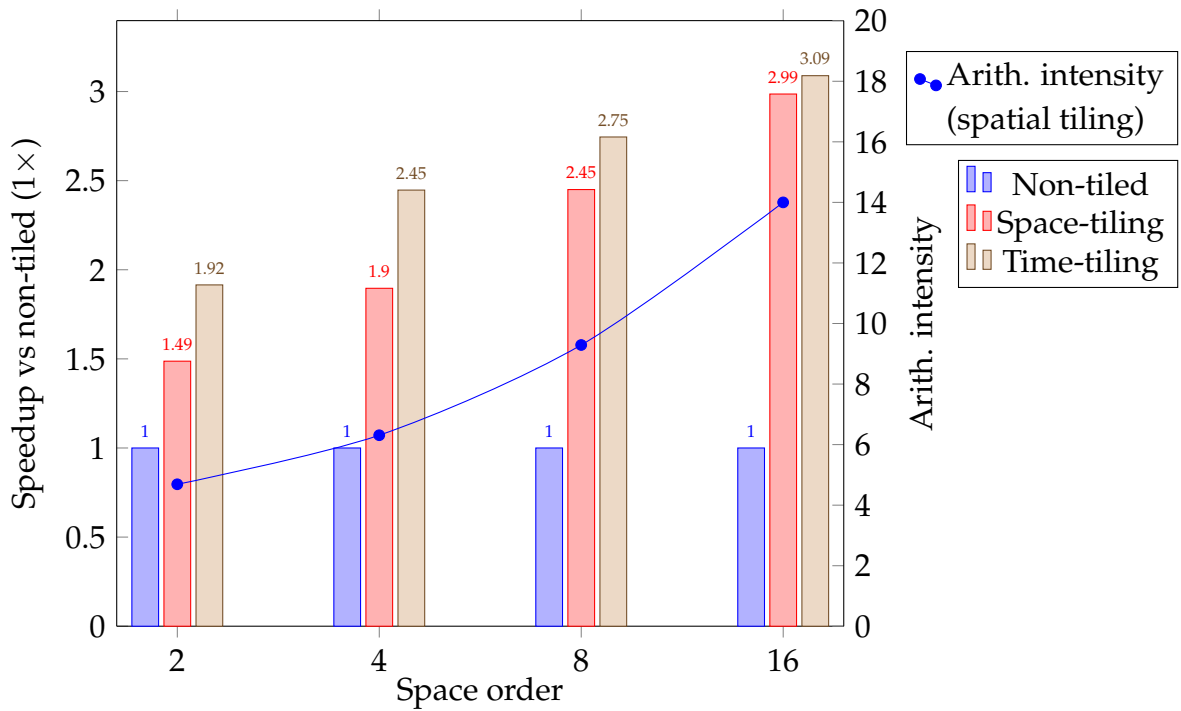
\begin{figure}[p]
\centering
\begin{tikzpicture}
\begin{axis}[
width=0.7\textwidth,
legend style={at={(1.46,0.77)},anchor=north east},
xlabel={Space order},
xtick distance=1,
symbolic x coords={2,4,8,16},
nodes near coords,
axis y line*=left,
ylabel near ticks, yticklabel pos=right,
ylabel={Speedup vs non-tiled (\(1\times\))},
ymin=0,
ybar,
every node near coord/.append style={font=\fontsize{6.5}{6.5}},
]
\addplot table [x=sp,y=run,col sep=comma] {data/laplace-nt600.csv};
\addplot table [x=sp,y=run,col sep=comma] {data/laplace-sp600.csv};
\addplot table [x=sp,y=run,col sep=comma] {data/laplace-tm600.csv};
\legend{Non-tiled,Space-tiling,Time-tiling}
\end{axis}

\begin{axis}[
width=0.7\textwidth,
legend style={at={(1.46,0.95)},anchor=north east,cells={align=left}},
axis y line*=right,
ylabel near ticks, yticklabel pos=right,
axis x line=none,
symbolic x coords={2,4,8,16},
ymin=0, ymax=20,
ylabel={Arith. intensity},
ybar,
]
\addplot[smooth,mark=*,blue]
coordinates{
	(2,4.69)
	(4,6.31)
	(8,9.29)
	(16,14.0)
};
\legend{Arith. intensity\\(spatial tiling)}
\end{axis}
\end{tikzpicture}
\caption{Spatially and time-tiled runtimes for the Laplace operator compared with non-tiled, with a grid size of \(16 \times 600^3\) as above. The runtimes under tiling converge as arithmetic intensity increases.}
\label{fig:laplace-graph}
\end{figure}

As a preliminary, we note that both spatial and time-tiling yield much lower runtimes than non-tiled code.
This shows that non-tiled code is much less efficient than tiled code, highlighting the importance of the tiling transformation and data locality.

Immediately evident is that the benefit from time-tiling rapidly tapers as the space-order increases, with stencils with larger radii.
There are two likely (related) explanations for this: an increase in the amount of data needed to compute any one tile, and an increase in arithmetic intensity, slowing the rate of data consumption, making memory bandwidth less relevant.

\begin{table}[p]
\centering
\begin{tabular}{cr|c|c|ccccc}
\toprule
& & N & S & \multicolumn{5}{c}{T; time tile size} \\
Space-order & &   &   & 1 & 2 & 4 & 8 & 16 \\
\midrule
16 & \footnotesize GFlop/s & 31.22 & 93.24 & 94.19 & 95.34 & 96.44 & 93.63 & 94.14 \\
\multicolumn{2}{r|}{\footnotesize arith. intensity} & 15.8 & 14.0 & 14.0 & 14.9 & 15.5 & 15.7 & 15.9 \\
\midrule
8 & \footnotesize GFlop/s & 30.22 & 74.06 & 75.89 & 78.88 & 81.40 & 82.84 & 82.97 \\
\multicolumn{2}{r|}{\footnotesize arith. intensity} & 9.29 & 9.29 & 9.29 & 17.5 & 24.1 & 18.2 & 33.6 \\
\midrule
4 & \footnotesize GFlop/s & 28.96 & 54.91 & 55.98 & 61.92 & 65.35 & 66.44 & 70.85 \\
\multicolumn{2}{r|}{\footnotesize arith. intensity} & 6.31 & 6.31 & 6.31 & 12.6 & 23.7 & 23.8 & 24.7 \\
\midrule
2 & \footnotesize GFlop/s & 26.55 & 39.49 & 40.18 & 44.54 & 47.60 & 50.84 & 50.80 \\
\multicolumn{2}{r|}{\footnotesize arith. intensity} & 4.69 & 4.69 & 4.69 & 9.38 & 18.8 & 33.2 & 48.6 \\
\bottomrule
\end{tabular}
\caption{Performance and arithmetic intensity of stencils with non-tiled (N), spatially tiled (S), and time-tiled (T) computations and time tile sizes 1, 2, 4, 8, 16. Tile sizes used to calculate arithmetic intensity available in project archive.}
\label{tab:laplace-roofline}
\end{table}

\begin{figure}[p]
\centering
\begin{tikzpicture}
\begin{axis}[
width=0.7\textwidth,
legend style={at={(1.05,0.95)},anchor=north west},
xlabel={Arithmetic intensity},
ylabel={Performance (GFlop/s)},
xmode=log,
log basis x={2},
ymode=log,
log basis y={2},
log ticks with fixed point,
cycle list name=exotic,
]
\addplot table [x=oi,y=gflops,col sep=comma] {data/laplace-roof-so16.csv};
\addplot table [x=oi,y=gflops,col sep=comma] {data/laplace-roof-so8.csv};
\addplot[mark=triangle] table [x=oi,y=gflops,col sep=comma] {data/laplace-roof-so4.csv};
\addplot table [x=oi,y=gflops,col sep=comma] {data/laplace-roof-so2.csv};
\legend{Time-tiling so16,Time-tiling so8,Time-tiling so4,Time-tiling so2}
\addplot [domain=2:15.15,name path=A] {17.3*x};
\addplot [domain=15.15:50] {262.01};
\end{axis}
\end{tikzpicture}
\caption{Graph of performance against arithmetic intensity estimates for the Laplace operator, grid size \(512^3\) and 16 time iterations. Points (approximately) bottom-to-top represent increasing time tile sizes as in table above. The solid line is the maximum predicted by the roofline model and benchmarks; roofline truncated to avoid scaling. Observe that arithmetic intensity estimates are far tighter at higher space orders.}
\label{fig:laplace-roofline}
\end{figure}
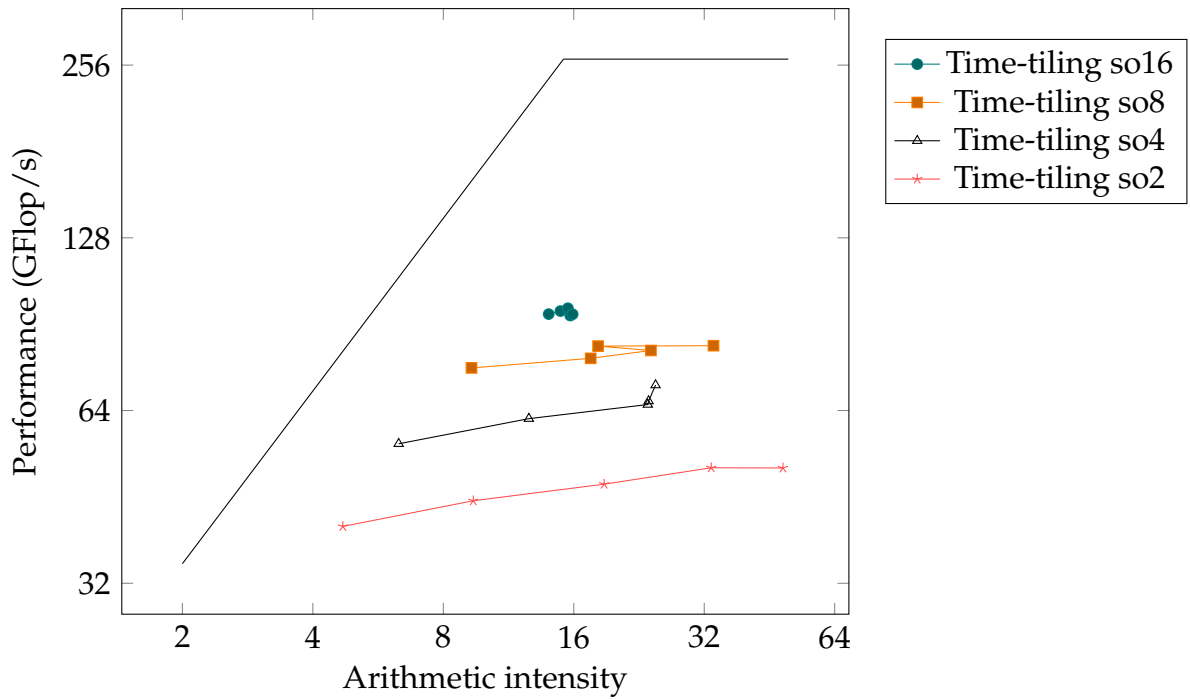

Table~\ref{tab:laplace-roofline} gives the arithmetic intensity and Devito-reported performance for each data point.
Note that arithmetic intensity increases significantly from a space-order of 2 to 16.
At the same time, reported GFlop/s of spatially tiled code increase dramatically from 15\% of that achieved by the \emph{LINPACK} benchmark, to over 35\%.
Therefore, we expect that we are nearing the computational bound of our test machine, and in accordance with the roofline model, would expect little improvement from memory bandwidth improvements such as time-tiling.

Figure~\ref{fig:laplace-roofline} is the corresponding plot to Table~\ref{tab:laplace-roofline}.
Note that we have only plotted the time-tiled results, as time-tiling with a time tile size of 1 is equivalent to spatial tiling and the results are nearly indistinguishable.
As predicted by the roofline model, GFlop/s achieved increases more for stencils of lower arithmetic intensity; observe that for space order 16, GFlop/s achieved is nearly constant, indeed decreasing for time tile sizes 8 and 16.

We can observe a quirk of the bounding scheme employed, as the arithmetic intensity does not always increase with an increase of the time tile size, which would ostensibly increase data transfer.
Indeed, there are two factors at work here: firstly, the sizes of spatial tiles chosen are not invariant when increasing time tile size.
Secondly, the bounding algorithm depends on a discrete condition being fulfilled, i.e.~\(F_i \ge \Omega\), but does not account for cache usage when it is close to this bound.

We see that it has had the greatest effect on the samples with space order 16; here it showed that the necessary data transfer for time-tiled computation was very close to that of spatially tiled computation.
This clearly predicts a minimal performance gain (obtained 3.3\%), as arithmetic intensity has not changed significantly.

Conversely, we may hypothesise that the arithmetic intensity of the lower space order stencils (such as space order 2) has not increased by as much as the estimator gives, as the increase in arithmetic intensity (\(10.4\times\)) has not been matched by a commensurate increase in the performance (\(1.22\times\)), although note that this would be determined partially by the memory bandwidth.

\section{Performance of the acoustic wave equation operator}
\label{sec:perf-awe}

\subsection{Application of the operator}
The acoustic wave equation (AWE) determines the propagation of acoustic waves, describing velocity as a function of space and time.
In our evaluation, we again used three spatial dimensions and a time dimension, with deterministically generated input.

In this section, the family of stencils generated by Devito are the acoustic wave equation operators governing wave propagation only,\footnote{Not including sources and receivers, required for some applications.} with space orders 4, 6, 8, 12, and 16.
Importantly for our evaluation, stencils generated by acoustic wave equation operator have lower arithmetic intensity than those from the Laplace operator, allowing us to distinguish spatially and time-tiled results more easily.

\subsection{Results}

Again, the grid size of \(512^3\) and a time buffer of 17 time iterations was chosen primarily as it was the largest grid that fulfilled our memory constraint.
Hence numerical verification was only performed for the last 17 iterations; nevertheless these iterations would represent the largest possible divergence of values, providing a reliable check of validity.
A buffer of at least 17 time iterations was necessary for the auto-tuner.

\begin{table}[p]
\centering
\begin{tabular}{rr|ccccc}
\toprule
& Grid size & \multicolumn{5}{c}{\( 512^3 \)} \\
& Space-order & 4 & 6 & 8 & 12 & 16 \\
\midrule
N & \footnotesize runtime (s) & 7.205 & 8.560 & 9.898 & 12.981 & 16.433 \\
S & \footnotesize runtime (s) & 4.558 & 4.595 & 4.655 & 4.892 & 5.608 \\
T & \footnotesize runtime (s) & 2.492 & 2.688 & 2.941 & 3.660 & 4.468 \\
& \footnotesize decrease (\%) & 45.3\% & 41.5\% & 36.8\% & 25.2\% & 20.3\% \\
\midrule
(t,8) & \footnotesize runtime (s) & * & * & 3.601 & 3.939 & 4.468 \\
(t,6) & \footnotesize runtime (s) & * & * & 3.348 & 3.660 & - \\
(t,4) & \footnotesize runtime (s) & 2.857 & 2.911 & 2.941 & - & - \\
(t,3) & \footnotesize runtime (s) & 2.688 & 2.688 & - & - & - \\
(t,2) & \footnotesize runtime (s) & 2.492 & - & - & - & - \\
\bottomrule
\end{tabular}
\caption{Runtimes from using the acoustic wave equation operator, space orders 4, 6, 8, 12, 16. N, S, T represent minimum runtimes without tiling, with spatial tiling, and with time-tiling respectively. (t,\(k\)) indicates results for time-tiling with a skewing factor of \(k\); T indicates the minimum of these. `-' and `*' denote invalid skewing factor and not tested respectively. Decrease is from spatial tiling to time-tiling.}
\label{tab:awe-results}
\end{table}

\begin{figure}[p]
\centering
\begin{tikzpicture}
\begin{axis}[
width=0.7\textwidth,
legend style={at={(1.46,0.77)},anchor=north east},
xlabel={Space order},
xtick distance=1,
symbolic x coords={4,6,8,12,16},
nodes near coords,
axis y line*=left,
ylabel near ticks, yticklabel pos=right,
ylabel={Speedup vs non-tiled (\(1\times\))},
ymin=0,
ybar,
every node near coord/.append style={font=\fontsize{6.5}{6.5}},
]
\addplot table [x=sp,y=run,col sep=comma] {data/awe-nt512.csv};
\addplot table [x=sp,y=run,col sep=comma] {data/awe-sp512.csv};
\addplot table [x=sp,y=run,col sep=comma] {data/awe-tm512.csv};
\legend{Non-tiled,Space-tiling,Time-tiling}
\end{axis}

\begin{axis}[
width=0.7\textwidth,
legend style={at={(1.46,0.95)},anchor=north east,cells={align=left}},
axis y line*=right,
ylabel near ticks, yticklabel pos=right,
axis x line=none,
ylabel={Arith. intensity},
ybar,
symbolic x coords={4,6,8,12,16},
ymin=0, ymax=8,
]
\addplot[smooth,mark=*,blue]
coordinates{
	(4,2.1)
	(6,2.5)
	(8,2.89)
	(12,3.36)
	(16,4.42)
};
\legend{Arith. intensity\\(spatial tiling)}
\end{axis}
\end{tikzpicture}
\caption{Spatially and time-tiled runtimes for the acoustic wave equation operator compared with non-tiled, with a grid size of \(512^3\) as above. Time-tiling runs in 45\% less time than spatial tiling, decreasing to 20\% as arithmetic intensity increases.}
\label{fig:awe-runtime}
\end{figure}
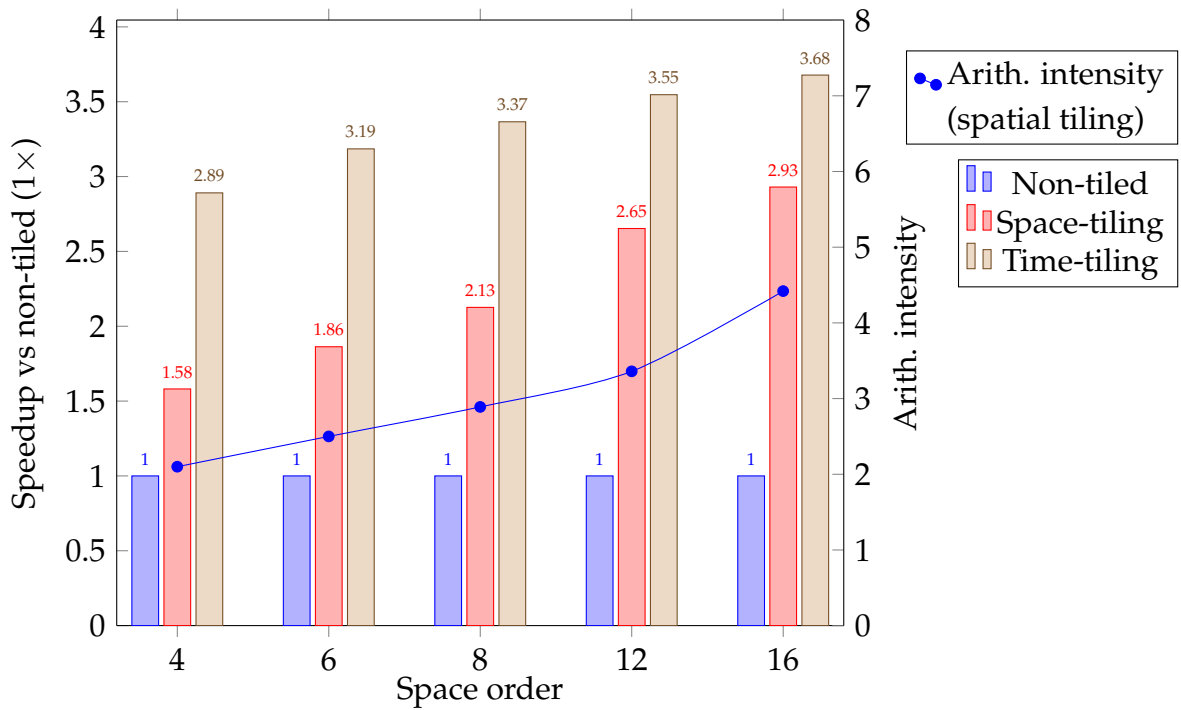

In Table~\ref{tab:awe-results}, we again observe the greatest decrease in runtime (45\%) at the lowest space orders, tapering off as the space order increases, although a minimal gain of 20\% is still had for the largest space order tested.
There is a large step in runtime decrease (36.8\% to 25.2\%) between space orders 8 and 12, whereas the steps between each of the other space orders are less than 5\%.
Note that we have excluded some combinations from testing based on observations from the Laplace operator and the other combinations with this operator, which we discuss in Section~\ref{sec:eval-skewing-effect}.

Figure~\ref{fig:awe-runtime} gives the corresponding graph; we are able to make the same observations as for the Laplace stencils.
However, with the acoustic wave equation, we observe that time-tiling always gives a significant decrease in runtime over spatial tiling, this is likely due to the lower arithmetic intensity of these stencils.

In all cases, the minimum runtime was produced with the smallest valid skewing factor, even if it was not a power of two.
Additionally, all of the minimum runtimes occurred with the auto-tuner choosing a time tile size of 16.

\begin{table}[p]
\centering
\begin{tabular}{cr|c|c|ccccc}
\toprule
& & N & S & \multicolumn{5}{c}{T; time tile size} \\
Space-order & &   &   & 1 & 2 & 4 & 8 & 16 \\
\midrule
16 & \footnotesize GFlop/s & 16.07 & 47.10 & 47.52 & 54.38 & 58.06 & 57.75 & 59.11 \\
\multicolumn{2}{r|}{\footnotesize arith. intensity} & 4.42 & 4.42 & 4.42 & 4.89 & 4.64 & 4.53 & 4.47 \\
\midrule
12 & \footnotesize GFlop/s & 16.86 & 44.75 & 44.38 & 52.43 & 57.99 & 57.98 & 59.82 \\
\multicolumn{2}{r|}{\footnotesize arith. intensity} & 3.36 & 3.36 & 3.36 & 5.82 & 4.79 & 4.63 & 4.55 \\
\midrule
8 & \footnotesize GFlop/s & 17.46 & 37.12 & 36.96 & 47.21 & 54.46 & 55.74 & 58.76 \\
\multicolumn{2}{r|}{\footnotesize arith. intensity} & 2.89 & 2.89 & 2.89 & 5.78 & 8.23 & 6.07 & 5.92 \\
\midrule
6 & \footnotesize GFlop/s & 17.45 & 32.51 & 32.39 & 42.65 & 50.40 & 53.11 & 55.57 \\
\multicolumn{2}{r|}{\footnotesize arith. intensity} & 2.50 & 2.50 & 2.50 & 5.00 & 8.67 & 7.12 & 6.89 \\
\midrule
4 & \footnotesize GFlop/s & 17.44 & 27.57 & 27.49 & 37.52 & 45.12 & 48.17 & 50.44 \\
\multicolumn{2}{r|}{\footnotesize arith. intensity} & 2.10 & 2.10 & 2.10 & 4.20 & 8.40 & 9.29 & 8.82 \\
\bottomrule
\end{tabular}
\caption{Performance and arithmetic intensity of stencils with non-tiled (N), spatially tiled (S), and time-tiled (S) computations and time tile sizes 1, 2, 4, 8, 16. Tile sizes used to calculate arithmetic intensity available in project archive.3}
\label{tab:awe-roofline}
\end{table}

\begin{figure}[p]
\centering
\begin{tikzpicture}
\begin{axis}[
width=0.7\textwidth,
legend style={at={(1.05,0.95)},anchor=north west},
xlabel={Arithmetic intensity},
ylabel={Performance (GFlop/s)},
xmode=log,
log basis x={2},
ymode=log,
log basis y={8},
log ticks with fixed point,
cycle list name=exotic,
]
\addplot table [x=oi,y=gflops,col sep=comma] {data/awe-roof-so16.csv};
\addplot table [x=oi,y=gflops,col sep=comma] {data/awe-roof-so12.csv};
\addplot[mark=triangle] table [x=oi,y=gflops,col sep=comma] {data/awe-roof-so8.csv};
\addplot table [x=oi,y=gflops,col sep=comma] {data/awe-roof-so6.csv};
\addplot table [x=oi,y=gflops,col sep=comma] {data/awe-roof-so4.csv};

\addplot [domain=1.5:4] {17.3*x};
\addplot [domain=2:5,dashed] {13*x};
\legend{Time-tiling so16,Time-tiling so12, Time-tiling so8,Time-tiling so6, Time-tiling so4}
\end{axis}
\end{tikzpicture}
\caption{Graph of performance against arithmetic intensity estimates for the acoustic wave equation operator, grid size \(512^3\) and 250 time iterations. Points (approximately) bottom-to-top represent increasing time tile sizes as in table above. The solid line is the maximum predicted by the roofline model and benchmarks; roofline truncated to avoid scaling. We see a convergence of arithmetic intensity and performance as time tile size increases.}
\label{fig:awe-roofline}
\end{figure}
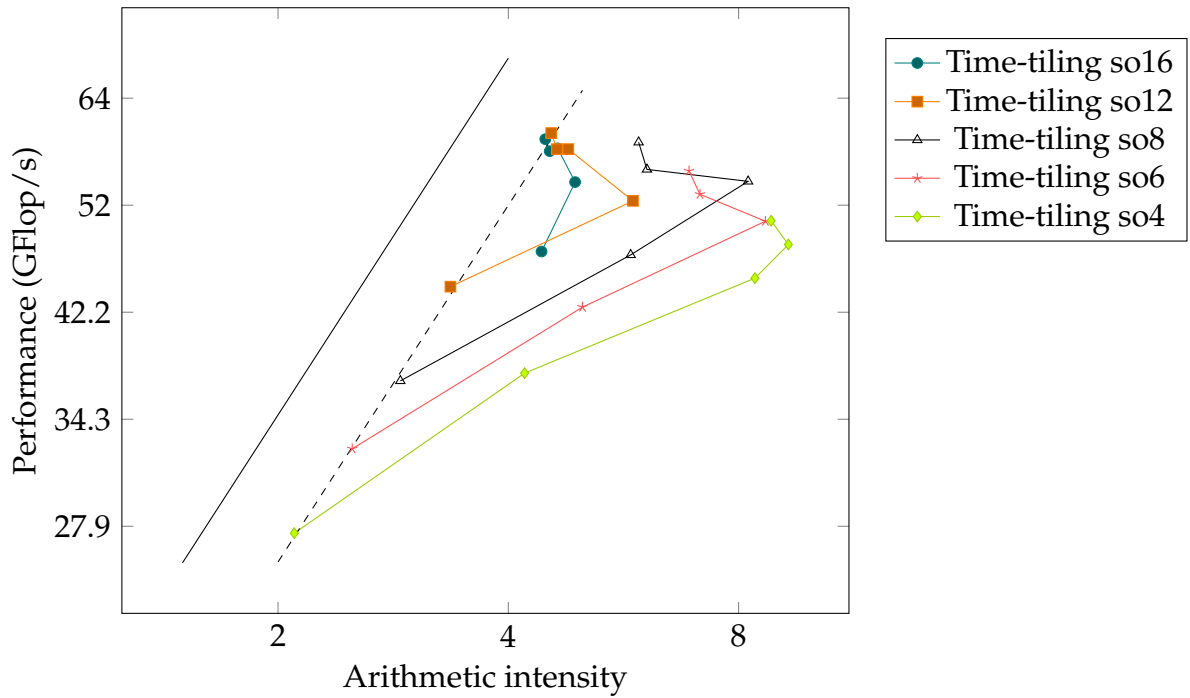

Table~\ref{tab:awe-roofline} gives the arithmetic intensity and Devito-reported performance for stencils of varying space order time tile size.
To gather these results, time tile size was fixed and the auto-tuner was used to search for the best choice of spatial tile sizes.
This is to produce an analysis of the arithmetic intensity under time-tiling and performance gains.

Figure~\ref{fig:awe-roofline} is the corresponding roofline plot from the table.
We observe that at lower space orders, we experience performance gains through to time tile sizes of 16, although for space orders 12 and 16, performance stagnates after time tile size 4.
We commented on the quirk of the bounding algorithm which gives rise to larger time tile sizes sometimes having lower arithmetic intensity, although the effect is perhaps more extreme here.

An interesting observation is that the points for space orders 12 and 16 have converged for time tile sizes 4 and larger, suggesting a limit of sorts.
We add a line (slope 13) that approximately intersects the points corresponding to tile size 1, which coincides with this cluster of points.
This suggests that the obtained memory bandwidth is around 13 GB/s, compared to 17.3 GB/s reported by the \emph{STREAM} benchmark.
Like the previous section, it is reasonable to expect that tighter bounds are possible for the arithmetic intensity under time-tiling.

Finally, we observe that the maximum achieved performance for any of these stencils is less than 60 GFlop/s, compared to over 90 GFlop/s using the Laplace operator.
A possible reason for this is the lower inherent arithmetic intensity of the acoustic wave equation stencils, and that we are approaching the performance ceiling for this operator and our set of optimisations.

\section{Effect of the skewing factor}
\label{sec:eval-skewing-effect}

\paragraph{Arithmetic intensity}
We do not propose changing the measure of arithmetic intensity of a stencil computation when skewing, as the same data is required for each computation, and the ordering of computations has not changed.\footnote{For a more concrete understanding of this, one may appeal to the polyhedral model which is beyond the scope of this work.}
Nevertheless, due to data non-alignment with odd skewing factors, skewing \emph{may} have the impact of reducing the effective arithmetic intensity, especially if there is a memory bandwidth bottleneck.
We believe that this is overwhelmingly offset by the increase due to time-tiling.\footnote{The reader observes that time-tiling is intended to increase (minimally double) the arithmetic intensity, and a data alignment issue will not reduce the arithmetic intensity by more than half. Additionally, the nature of the stencil means that data neighbouring a dependency would often have to be loaded in any case, to calculate neighbouring points.}

Empirically, observe that in all but one case,\footnote{Laplace operator stencil of space order 8 with grid size \(32 \times 500^3\) needed 3.762s with skew factor 8 and 3.764s with skew factor 4. The difference appears to be minimal; for comparison see the same stencil on the \(16 \times 600^3\) grid.} time-tiling with a smaller skewing factor produces a lower runtime than a larger skewing factor, and that there is a progressive increase in runtime as the skewing factor increases.
This held for our test machine even if the skewing factor was not a power of two, such as 3 or 6, which were compared against skewing factors of 4 and 8 respectively.

However, it is important to note that this may be architecture-dependent, and should be tested on other architectures rather than assumed.
Further, we have only tested against 10 members across two families of stencils, which may not be fully representative of stencils in general.

\section{Effect of the time tile size}
\label{sec:tt-size-effect}

Note that time-tiling with a time tile size of 1 is an analogue of spatial tiling, as each time iteration occurs with a new iteration of the time tile loop, and the incremental time loop executes exactly once when it is encountered.

With the acoustic wave equation operator (Table~\ref{tab:awe-roofline}), we notice that stencils with low space order benefit from increases in tile size up to 16 and possibly beyond, while those with higher space orders have little performance gains after a time tile size of 4 or 8.

In order to explain this, we consider the arithmetic intensity estimator which we introduced in Section~\ref{sec:arithmetic-intensity}.
We previously noted for both the Laplace and acoustic wave equation operators, bounding from our arithmetic intensity operator had a larger effect with larger time tiles.
We know that even with such a large bounding effect, we still overestimate the true arithmetic intensity of a stencil.

For the stencils with large space order, we believe that an increase in the time tile size provides no additional benefit beyond a point, as performance is once again bounded by memory bandwidth; as much of the time tile needs to be loaded from memory more than once, increasing the time tile size cannot reduce the memory bandwidth requirements.
Hence the performance of the stencil converges, and other techniques increasing the arithmetic intensity will be needed to yield performance gains.

\section{Evaluation of the arithmetic intensity estimator}
\label{sec:ai-eval}

\subsection{Conditions for most accurate estimates}
The rather counter-intuitive quirk of the arithmetic intensity estimator, that of larger time tile sizes sometimes resulting in lower arithmetic intensity, is due to a failure to account for the effects of moderate increases in time tile size.
The bounding effect is the greatest with large increases in time tile size, and this explains the decrease of the bound with larger time tile sizes.

One may observe that the problems surveyed were all in 3 spatial dimensions.
If this were to increase, keeping spatial dimension extents equal, we would likely observe a far tighter bound on arithmetic intensity using our method.
This is caused by an increase in the ratio of tile-face boundary volume to cache size; our estimator is most effective when this is maximised.

\subsection{Modelling performance improvement with the estimator}
Using the roofline model, one would expect the greatest performance improvement with the largest increase in estimated arithmetic intensity.
However, we observed that our estimated arithmetic intensity peaked before the greatest runtime decreases were had, even if the increases were gradual by that point.

We previously observed that our estimator fails to account for some memory traffic at smaller time tile sizes, and believe that this is the main cause of the inaccuracy.

The estimator, however, provides a good insight into when an increase in time tile size will \emph{not} provide a decrease in runtime, which as the values for arithmetic intensity converge for an arbitrary stencil.
This is nonetheless helpful in modelling performance improvement.

\subsection{Comparison to the naive estimator}
Nevertheless, the estimator used clearly predicts the convergence of performance in GFlop/s as the time tile size increases.
This is in contrast to the naive estimator, which scales proportionally to the arithmetic intensity, leading to extremely large figures for arithmetic intensity.
Such a biased estimator could lead one to conclude that no more performance gain could be had from an increase in memory bandwidth, as the point would be in the computationally limited regime (arithmetic intensity \(> 15.15\) for our test machine).
It is evident, then, that our estimator is far more useful in this regard.

Note, however, that the naive estimator works adequately in the case of spatial tiling, as the cumulative tile-face boundaries tend to be smaller than or only slightly more than the cache size.

\subsection{Constructing a better estimator}
It should be possible to construct an estimator which accounts for gradual increases of the tile face boundaries, particularly as the boundary size just begins to exceed the cache size.
Such an estimator would remove the `edge' encountered as some \(F_i > \Omega\) by a small amount, reducing the overestimate for smaller time tile sizes.

\section{Limitations and further evaluation}
\label{sec:further-eval}

We were careful to guarantee the validity of our specific results, as documented in our testing methodology; this section illustrates how stronger conclusions may be drawn if the limitations were surmounted.

\subsection{Realism of test problems}

\subsubsection{Source and receiver loops}
Previously seen in Section~\ref{sec:impl-imperfect}, our implementation of time-tiling is unable to handle imperfectly-nested loops, source and receiver loops in particular.
As these are an important case for Devito to optimise, it is necessary to implement support for source and receiver loops and validate the results from our performance testing without these loops.

\subsubsection{Operators}
We have analysed stencils from two families of operators; one family, the acoustic wave equation, is extremely important in Devito's original target domain,
making it especially relevant.
However, there are other closely-related families of stencils from wave equations that are also highly relevant to Devito's use cases, and it would be beneficial to perform a comparative evaluation between them.

\subsubsection{Problem sizes}
We were limited to 64GB of memory on our test machine.
As we had to choose large enough spatial dimensions that each time iteration would dwarf the processor's cache, we were restricted to spatial dimensions of \(500 \times 500 \times 500 \) and larger.
This in turn restricted the number of time iterations we could store, even using time-buffering.
Indeed, the auto-tuner required at least 17 iterations to test time tile sizes up to 16, we were effectively prevented from testing time tile sizes of 32 (requiring 33 iterations of storage), as additional space was required for numerical verification.

Therefore, we were unable to perform analysis with both larger problem domains, or larger time tile sizes, which would have enabled a more complete evaluation.

\subsection{Test architecture}
Our testing was completed on a single-socket machine with one 8-core Sandy Bridge Xeon processor, and all tests were run with 16 threads, which it supports.
This provides a limited perspective on how time-tiling may perform on other systems and architectures, particularly those with more cores, cache, or distributed memory.

\subsection{Memory analyses}
For a more in depth study, memory analyses measuring cache misses and memory traffic should be performed, to investigate how time-tiling has improved these.
Additionally, such analyses would reveal more information on the true arithmetic intensity of the computation.
This would allow us to determine how the bottlenecks have changed after time-tiling, how far the performance is from theoretical limits, and reveal further areas of optimisation.

Recording cache misses and precise memory traffic data would also allow for the assessment of our novel arithmetic intensity operator, and provide hints on possible improvements.

\subsection{Comparison to spatial tiling}
In our analysis, we have compared time-tiling to spatial tiling code, both of which used our new tiling algorithm which utilised min/max bounding.
It is possible that this is slower for spatial tiling than the previous remainder loop tiling algorithm in Devito.

We have decided to accept this possibility, as any difference is unlikely to be significant; remainder loops incur a synchronisation penalty, whereas min/max bounding requires a few additional arithmetic operations.
Neither of these should take up a significant amount of time compared to the time for the stencil computation.

As a note, it may be possible to reduce the bounding overhead slightly by avoiding the floating-point \texttt{fmin} and \texttt{fmax} functions in bounding, as only integer comparisons are needed.
This is documented in Section~\ref{sec:impl-minor}.

\subsection{Time-tiling with larger time dependences}
All of the stencils surveyed only depended on data from the previous time iteration (\(\delta_t=1\)).
In some real-world applications, a larger time dependence is used through multi-step linear methods, such as the Adams-Bashforth method.

It is worth understanding the effect of time-tiling on such applications, as an intuitive assessment of arithmetic intensity suggests that the performance gains when \(\delta_t > 1\) will be even larger than the current scenario.
To see this, realise that under spatial tiling, each time iteration would require twice as much data to be loaded from memory, but under time-tiling, this is analogous to taking a time tile size of 2, with much smaller memory transfer required.

Clearly, this is a hypothesis rather than a rigorous analysis of such a stencil, but it certainly demands further scrutiny.

\subsection{Further uses for an arithmetic intensity estimator}
The Devito auto-tuner currently attempts a wide variety of combinations in its search space, even implausible ones with excessively large tile sizes, which do not represent significant improvement over non-tiled code.
This is inefficient, as these auto-tuner trials take the longest time to run, potentially dominating the runtime of auto-tuning.

A bound assuming perfect scheduling could be used to prune poor tile size combinations from the auto-tuner search space, decreasing auto-tuning time.
Using a bounding approach is also easily extensible to other architectures, so this could be implemented in a straightforward manner.

Finally, the algorithm used to construct the bound may provide hints as to good tile size combinations, for example that choosing \(t_1 \ge t_2\) often leads to decreased runtime.
However, this heuristic should not be weighted too heavily, as the lower levels of cache and architectural differences will play a role in determining runtime.

\section{Summary}
\label{sec:eval-conclusion}

In this chapter we presented a robust testing methodology and model to understand the performance increase from time-tiling, and when the decrease in runtime is maximal.
We used the roofline model to show that the stencil computations were still bounded by memory bandwidth bottlenecks.

Through our investigation, we discovered a runtime decrease of up to 45\% in an extreme case and at least 20\% in other cases using multiple real-world stencils from the family of acoustic wave equation operators.

We also analysed the effect of the parameters presented by time-tiling, the skewing factor and time tile size, on runtime of a stencil computation, and identified when the resultant runtime decrease was maximal.

We additionally presented an estimator for arithmetic intensity under time-tiling, establishing its bounds and limitations, and use it to successfully model the gradual decrease in performance benefits of our time-tiling transformation.

Finally, we critique our evaluation, highlighting its shortcomings and areas for deeper evaluation, as well as how our analysis could be applied in Devito.

\paragraph{A final remark}
Previous analysis of the stencil of space order 8 with the acoustic wave equation operator yielded a decrease in runtime of 27.5\%, when passing non-tiled code from Devito to the polyhedral compiler CLooG~\cite{dylan}.
We surpassed this, demonstrating a 36.8\% decrease in runtime for the same stencil with our native time-tiling algorithm in Devito, albeit with the power of Devito auto-tuner and a larger time tile size.

	\documentclass[thesis.tex]{subfile}

\chapter{Conclusion}
\label{ch:conclusion}

\section{Context and review}

This project drew together aspects of compiler transformations, differential equations and computation, computer architecture, and software engineering.
The different perspectives of the fields provided immense insight into the complexity of numerically approximating partial differential equations---a far cry from anything Taylor may have imagined in the 18th century!
These insights culminated in a transformation yielding significant performance gains in Devito's target domain: seismic imaging.
Though time-tiling is a well-established optimisation, this implementation is a necessary and important step for Devito.

During our evaluation in Chapter~\ref{ch:evaluation}, we used Devito's abstractions to effortlessly generate entire families of stencils by varying the space order of a differential equation; producing these stencils from the equations is decidedly non-trivial, and would have otherwise been extremely time-consuming.
As a consequence, this project includes a broad survey of time-tiling on stencil kernels from real-world applications.

While the concept of time-tiling is superficially straightforward, its implementation was decidedly not so: unlike interfacing with a `back-end' polyhedral compiler, such as CLooG or Pochoir (Section~\ref{sec:pochoir}) with all its attendant integration problems, implementation within Devito meant understanding the numerous stages of internal representation used to transform a differential equation into code.
Without these layers, Devito would not be able to expose APIs at different levels for more intricate applications; making a transformation that would work with these was a major hurdle.

The performance evaluation was enlightening and challenging in its own ways.
It is easy to be convinced of the benefits of time-tiling from the literature, and another experience altogether to witness real decreases in runtime while watching a wall clock;
mustering the scepticism required to confirm its validity required much deeper reference to computer architecture and computation than implementing the transformation had needed.

\section{Contributions}
A summary of the contributions of this project.

\begin{itemize}
	\item The headline contribution of the project; documentation and implementation of the time-tiling transformation for perfect loop nests in Devito, with accompanying test cases and auto-tuner enhancements.
	This is a necessary and significant step toward implementing time-tiling for Devito in its full generality, including tiling of imperfect loop nests.

	\item We outlined the remaining work in implementation, including analyses of their importance and difficulty, and their consistency with our implemented transformation.

	\item We presented and critiqued our testing methodology and models, which we used for the performance evaluation of our implemented time-tiling transformation.
	In particular, we devoted attention to numerical verification, the set-up of the test environment, and the modelling of performance.

	\item We demonstrated a runtime decrease of up to 45\%, and in general more than 20\%, with the transformation compared to Devito's existing optimisations on kernels with real-world applications, particularly the stencils of varying space orders arising from the acoustic wave equation, of great importance to Devito's target domain, seismic imaging.
	In so doing, we validated previous work which demonstrated a runtime reduction of up to 27.5\%~\cite{dylan}.

	\item We proposed an estimator for \emph{arithmetic intensity under time-tiling} which extends Devito's existing estimator, including a proof that it is consistent with the widely-cited roofline model.
	We determined the performance of the estimators under varying conditions, and used our estimator to demonstrate that the evaluated stencils were still bound by memory bandwidth.

	\item Further, we developed and critiqued a model estimating performance improvement based on our arithmetic intensity estimator, which is able to predict the circumstances under which minimal or no performance improvement is seen.

	\item We developed a hypothesis on how the skewing factor affects the runtime of a stencil, which holds in our test architecture, and in particular our experimentation.
\end{itemize}

\section{Future work}
\label{sec:future-work}

Apart from work in evaluation and extensions to time-tiling discussed below, we also like to highlight Section~\ref{sec:impl-minor} (`Minor extensions to time-tiling').

\subsection{Tiling imperfectly-nested loops}
In this work, we implemented the time-tiling transformation for perfect loop nests in Devito.
However, Devito is also used to solve problems that do not naturally generate perfect loop nests.
An example would be source and receiver loops, which may be used with the acoustic wave equation to model the propagation of waves to study unknown media.
We briefly discussed this additional transformation in Section~\ref{sec:impl-imperfect}.

\begin{figure}[!ht]
\begin{lstlisting}
for (int t = t_s; t < t_e; t++) {
  for (int x = x_s; x < x_e; x++)
    for (int y = y_s; y < y_e; y++)
      A[t][x][y] = A[t-1][x-1][y+1] + A[t-1][x+1][y-1];
  for (int j = j_s; j < j_e; j++)
    A[t][j][j+1] *= 1.5;
}
\end{lstlisting}
\caption{An imperfectly-nested loop. The body of the \texttt{t} loop contains two loops; in a perfect loop nest, only the body of the innermost loop can contain multiple statements, none of which can be loops.}
\label{lst:loop-imperfect}
\end{figure}

Figure~\ref{lst:loop-imperfect} gives an example of generated source and receiver loops in Devito.
Our tiling implementation is not able to time-tile this loop, although it may spatially tile the inner loop nest.

\subsubsection{Strategy for tiling imperfectly-nested loops}
In order to tile an imperfectly-nested loop, it must be transformed into a perfect loop nest.
To do this, transformations such as loop fission (previously discussed), loop fusion, and code sinking are used.
Ahmed et al.~generalise these transformations as conditionals in the body of the innermost loop when iterating over the \emph{product space}~\cite{ahmed-imperfect}, which is generated by the original variables of the imperfectly-nested loop; this generates a perfect loop nest of higher dimension, or a deeper loop nest.
Lim illustrates this particularly well using a slightly different strategy, affine partitioning~\cite{lim-affine-part}.

Again, the concepts for transforming imperfectly-nested loops to perfectly-nested loops are intuitive, but potentially tedious to generalise.
Ahmed explicitly provides an algorithm for doing so and analysis of the generated structure~\cite{ahmed-synth}; neither is remotely trivial.
In particular, evaluation should be performed to ascertain any runtime improvement between time-tiled stencils needing source and receiver loops and the same loops under spatial tiling.

\subsubsection{Consistency with time-tiling}
Manipulating a loop nest involves additional transformations.
It is clear from the literature that these are semantically compatible with tiling.
For skewing, consider that no change has been made to the execution order; loop index offsets are balanced by an equal and opposite change in loop variables.
Therefore, the necessary code transformations are not affected by skewing.

\subsection{Advanced manipulation of symbolic expressions}
\label{sec:future-aggressive}

The Devito symbolic engine (DSE) has additional modes which we have not discussed in detail in this report.
These enable it to perform more advanced transformations on the symbolic expressions that make up a differential equation.
In particular, they search for sub-expressions that should be assigned to temporaries, to avoid needless re-computation.

Our implementation separates skewing into a separate DSE mode, so that it will not occur alongside these advanced transformations.

\subsubsection{Consistency with time-tiling}
First, observe that these transformations were compatible with spatial tiling which was previously implemented using the remainder loop strategy.
Since it is legal to permute the spatial tiles within a single time iteration, our bounding strategy is guaranteed to be valid under spatial tiling.

For time-tiling, consider that our implementation does not parallelise the time dimension.\footnote{This is a consequence of our skewing factor being \emph{equal} to the largest spatial dependence distance. Increasing this by one would allow interchange of the incremental time loop with any increment spatial loop, however at least one incremental loop must not be parallelised.}
Thus the extraction will be valid, as all extracted code will be placed within the incremental time loop.

\subsubsection{Possible pitfalls in implementation}
This transformation extracts sub-expressions from \emph{SymPy} expressions, potentially changing the order of floating-point operations.
An obvious solution would be to perform skewing after this transformation, as we established in this work that skewing does not change the execution order.
A caveat is that skewing affects the behaviour of common sub-expression elimination (Section~\ref{sec:impl-cse}), which \emph{does} result in an execution order change in Devito.

This should be overcome with using a test case with suitable bounds; as Devito does not guarantee an order of operations on floats (due to optimisation), differences in values due to re-ordering are acceptable in principle.

A final remark is that skewing manifests in two parts: loop indices are skewed in the DSE, but the loop variable is not skewed, as the loop structure has not been generated yet.
Instead, the skewing factor is maintained as an attribute of the stencil until it is transformed into a loop.
This fact may explain how skewing and the other optimisations change the execution order, and it might be possible to surmount this change.

\subsection{Memory analyses to determine further areas for optimisation}
Due to time constraints, we did not perform memory analyses to determine cache misses and precise measures of memory traffic, instead focusing on bounding using arithmetic intensity.
While this has the advantage of being easily extended to other platforms and architectures, it would be useful to perform detailed analyses on our specific stencils.

This would reveal how bottlenecks have changed after time-tiling, particularly in the intermediate time tile sizes, for which our arithmetic intensity estimator was less effective.

Of special interest is how tiling and other transformations could increase the arithmetic intensity further, to bring stencils out of the memory-bound regime.
Additionally, validation and analysis of our techniques and models could lead to understanding of how similar techniques could be developed to study time-tiling on other architectures.

\subsection{Improvement of the arithmetic intensity estimator bounds, and uses thereof}
In connection with the previous section, memory analyses would be helpful to assess our assumptions regarding cache usage when constructing our arithmetic intensity estimator.
This would enable a more realistic model, although some sacrifices may be necessary with regard to overestimation.

As mentioned in Section~\ref{sec:further-eval}, an accurate arithmetic intensity estimator could be used to improve the optimisation process, such as by providing an auto-tuner heuristic which is easily extensible to other architectures.
This would enable detailed performance analysis without needing to resort to memory analyses, which are again not portable between architectures.

The main issue with the arithmetic intensity estimator stated in this work is its large bounds with moderate time tile sizes.
If a slightly more intricate estimator could be devised to deal with this scenario, its applicability would increase dramatically.

\subsection{Evaluation on more stencils and architectures}
Applying time-tiling to more families of stencils and other architectures is necessary to understand the transformation in greater depth.
A natural starting point would be other wave equations relevant to Devito, or distributed computing architectures, as evaluated by the OPS project.

In Section~\ref{sec:further-eval}, we discussed the relevance of multi-step linear methods to improve the accuracy of approximations, such as the Adams-Bashforth method.
These also apply to Devito's target domain, and should be considered in future evaluation work.
We proposed in the referenced section that time-tiling may yield greater reductions in runtime when using such methods.

Finally, with the implementation of time-tiling for imperfectly-nested loops as proposed above, it would be possible to evaluate the transformation in more generality, as it would support the tiling of source and receiver loops in the acoustic wave equation stencils, for example.
As this is an important use case for Devito, an implementation and evaluation of this functionality would certainly be an ambitious objective of considerable utility.

	\clearpage
	\addcontentsline{toc}{chapter}{Bibliography}
	\bibliographystyle{plain}
	\bibliography{thesis}
\end{document}